\documentclass[aps,prb,twocolumn,amsmath,amssymb,superscriptaddress,floatfix,scrartcl]{revtex4-1}

\usepackage{amssymb}
\usepackage{color}
\usepackage{graphicx}
\usepackage{mathrsfs}
\usepackage[unicode=true,pdfusetitle,
bookmarks=false,colorlinks=true,citecolor=blue,urlcolor=blue,linkcolor=red]{hyperref}
\usepackage{bm}
\usepackage{braket}
\begin{document}

\title{Holographic Duals of Inhomogeneous Systems: The Rainbow Chain and the Sine-Square Deformation Model}

\author{Ian MacCormack}
\affiliation{Kadanoff Center for Theoretical Physics, University of Chicago,
Chicago, Illinois 60637, USA}
\author{Aike Liu}
\affiliation{Department of Physics, University of Illinois at Urbana-Champaign,
Urbana, Illinois 61801, USA}
\author{Masahiro Nozaki}
\affiliation{Kadanoff Center for Theoretical Physics, University of Chicago,
Chicago, Illinois 60637, USA}
\author{Shinsei Ryu}
\affiliation{Kadanoff Center for Theoretical Physics, University of Chicago,
Chicago, Illinois 60637, USA}
\affiliation{James Franck Institute, University of Chicago, Chicago, Illinois 60637, USA}

\date{\today}

\begin{abstract}
  A holographic dual description of inhomogeneous systems is discussed.
  Notably, finite temperature results for the entanglement entropy in both the
  rainbow chain and the SSD model are obtained holographically by choosing
  appropriate foliations of the BTZ spacetime.
  Other inhomogeneous theories are also discussed.
  The entanglement entropy results are verified numerically,
  indicating that a wide variety of inhomogeneous field theory phenomenology can be seen in different slicings of asymptotically $AdS_3$ spacetimes.
\end{abstract}

%\pacs{72.10.-d,73.21.-b,73.50.Fq}

\maketitle
\tableofcontents
%\newpage

\section{Introduction}

Spatial inhomogeneity is ubiquitous in extended
quantum systems (quantum many-body systems)
realized, e.g., in solid states and cold atomic gases.
One source of inhomogeneity
is impurities or randomness,
but it can also be introduced intentionally,
for example via a harmonic trap confining cold atomic gas. 
%One typical form of inhomogeneity is impurities or randomness.
%In many cases, systems stay statistically translationally invariant.
%In other cases, such as cold atomic gases in a harmonic trap,
%this is not the case. 
  Inhomogeneity can have a dramatic effect in quantum many-body systems,
  e.g., it can alter the ground states completely -- we will discuss
  some examples momentarily.
  On the theoretical side,
  the vast majority of past work has focused on homogeneous systems,
  and we need to develop new tools to deal with such systems.
%  (Perhaps we need to elaborate on these points a little bit more..)
%Effects of inhomogeneity in quantum many-body systems have been studied.}
%Inhomogeneous quantum many-body systems (e.g., spin chains) have been studied recently.

In this paper, we study a series of inhomogeneous
quantum many-body systems in (1+1) dimensions constructed in the following way:
Let us start from a homogeneous quantum many-body Hamiltonian 
\begin{equation}
  H = \int dx\, \mathcal{H}(x)
\end{equation}
where $\mathcal{H}$ is the Hamiltonian density,
given by the $00$ component of the energy-momentum tensor,  
$\mathcal{H}\sim T_{00}$.
The spatial manifold here can be non-compact
(i.e., infinite line $x\in [-\infty, +\infty]$)
or
compact (i.e., circle of a finite circumference $L$, $x\in [0, L]$).
Similarly, for a homogeneous system 
defined on a one-dimensional lattice, 
one can consider the Hamiltonian
$H = \sum_i \mathcal{H}_{i}$,
where $\mathcal{H}_{i}$ is the Hamiltonian density
at a given lattice site $i$.
By ``deforming'' $H$
by an envelope function $f(x)$, 
we then consider an inhomogeneous system
\begin{align}
  H[f] = \int dx\, f(x)\, \mathcal{H}(x),
\end{align}
or $H[f] = \sum_i f(x_i) \mathcal{H}_{i}$ for lattice systems.

Of central focus in this paper are
the fundamental properties of the deformed Hamiltonians,
mainly the scaling of the entanglement entropy
both at zero and finite temperatures.
In this work, we will focus on the cases where
the original, homogeneous, Hamiltonians are  
those of (1+1)d conformal field theories (CFTs).

\subsection{Rainbow chains}

One deformation of central interest
is the so-called rainbow chain. 
\cite{2010NJPh...12k3049V,
  2014JSMTE..10..004R,
  2015JSMTE..06..002R,
  2016arXiv161108559R,
  2018arXiv180704179A,
  2018JSMTE..04.3105T}
The rainbow chain is an inhomogeneous (1+1)d quantum lattice model,
in which the envelope function $f(x)$ 
decays exponentially, 
\begin{align}
f(x) = e^{- h|x|},
\end{align}
where $h$ is a parameter.
For example, 
for the free fermion hopping model, we consider 
\begin{gather}
    {H}= \sum_{i,j=1}^N t_{ij} {c}^\dagger_i {c}^{\ }_j,
    \nonumber \\
    t_{ij}= -f_i\delta_{i-j,1} -f_i \delta_{i-j,- 1},
    \label{numerH}
\end{gather}
where
${c}^{\dag}_i$/${c}^{\ }_i$ are the fermion
creation/annihilation operators at site $i$,
$N$ is the total number of lattice sites, and
\begin{align}
f_j=  e^{  -h  |j -N/2| }
  \label{envelope rainbow}
\end{align}
(where the center has been shifted to $j=N/2$).
The homogeneous counterpart (where $f_j={\it const.}$)
realizes, 
at half-filling (partial filling),
the $c=1$ free fermion CFT in the continuum limit.

An interesting feature of this model  
is that the entanglement entropy shows volume law scaling;
It was found that
the entanglement entropy of the reduced density matrix,
when the chain is bipartitioned at the center
and the half of the chain is traced out,
grows linearly with respect to the subsystem size.
%$S_A(L_A)\simeq (c/6) h L_A$.
%
%
%the half-chain entanglement entropy
%of the ground state of the rainbow chain behaves as
%\begin{align}
%	S_A(L_A) 
%	=
%	\frac{c}{6} \log
%	\left(
%	\frac{ e^{h L_A} -1}{h} 
%	\right)
%	+ mb m
%  \cdots
%%	c'
%\end{align}
%for the sub region $A$ of our interest of length $L_A$. 
%Here, $c$ is the central charge of the relevant CFT.
%In the limit $h L_A \gg 1$, the 
%entanglement entropy grows extensively, 
%$S_A(L_A)\simeq (c/6) h L_A$.

It has been also understood that, in the continuum,
introducing the rainbow chain deformation is equivalent
to putting the CFT 
on a curved spacetime with the metric:
\cite{2016arXiv161108559R}
\begin{align}
  ds^2_{{\it AdS}_2}= -e^{ - 2h |x|} dt^2 + dx^2.
\end{align}
This is the metric of ${\it AdS}_2$.
Here, $h$ is the curvature scale (the inverse radius) of
${\it AdS}_2$.
By the change of the coordinates
\begin{align}
\eta = \mathrm{sgn}\, (x) \frac{e^{h|x|}}{h},
    \label{properLenCoordTransform}
\end{align}
the metric can also be written as
\begin{align}
  ds^2_{{\it AdS}_2}
  =
  \frac{1}{h^2 \eta^2}
  \left(
  d\eta^2 - dt^2
  \right),
  \label{ds AdS2}
\end{align}
which is the Poincar\'e patch of ${\it AdS}_2$.

%\begin{align}
%	H &=
%	-
%	\frac{t}{2}
%	\sum_{m=1/2}^{L-3/2} e^{-h m}
%	\left[
%	c^{\dag}_m c^{\ }_{m+1}
%	+
%	c^{\dag}_{-m} c^{\ }_{-(m+1)}
%	\right]
%	+ h.c. 
%      \nonumber \\
%  &\quad
%	-\frac{t}{2} c^{\dag}_{1/2} c^{\ }_{-1/2}.
%\end{align}

\subsection{M\"obius and SSD deformations}

Another interesting class of deformations
are M\"obius deformations and
the sine-square deformation (SSD).
\cite{2009PThPh.122..953G,
  2011PhRvB..83f0414H,
  2011PhRvA..83e2118G,
  2011PhRvB..84k5116S,
  2011PhRvB..84p5132M,
  2011JPhA...44y2001K,
  2012JPhA...45k5003K,
  2012PhRvB..86d1108H,
  2015JPhA...48R5208O,
  2015MPLA...3050092T,
  2015JPhA...48E5402I,
  2016arXiv160309543O,
  2016IJMPA..3150170I,
  2016IJMPA..3150170I,
  2016PhRvB..93w5119W,
  2017arXiv170906238T,
  2018PhRvB..97r4309W,
  2018arXiv180500031W,
  2018PTEP.2018f1B01T}
Starting from the uniform system defined on a spatial circle of
circumference $L$, the M\"obius evolution is given by
\begin{align}
  f(x)
  &= 1 - \tanh(2\gamma)
    \cos \frac{2\pi x}{L}.
%  \frac{1}{2}
%  \left(
%  e^{ +2\pi i x/L} + e^{-2\pi i x/L}
%  \right).
\end{align}
Here, $\gamma$ is a parameter;
$\gamma=0$ corresponds to the uniform
Hamiltonian, whereas when $\gamma\to \infty$,
\begin{align}
  \label{SSD envelope}
  f(x)
  %&= 1 -
  %\frac{1}{2}
  %\left(
  %e^{ +2\pi i x/L} + e^{-2\pi i x/L}
  %\right)
  %  \nonumber \\
  = 1 - \cos \frac{2\pi x}{L}
  %  \nonumber\\
  %&=
  %  \frac{-1}{2}
  %  \left(
  %  e^{ +2\pi i x/L} -2 + e^{-2\pi i x/L}
  %  \right)
  %  \nonumber \\
  =
    2 \sin^2 \frac{\pi x}{L}.
\end{align}
The resulting evolution operator
is called the sine-square deformation (SSD)
of the original Hamiltonian.
Correspondingly,
one can consider the lattice Hamiltonian
\eqref{numerH},
now with the hopping amplitude
$f_j= 2 \sin^2 \left(  {j \pi}/{N} \right)$.
 
The initial interest in the SSD
comes from the observation that
the ground state of the SSD Hamiltonian on an open chain
(when the system is described by CFT),
is equal to the ground state of
the uniform Hamiltonian on a finite circle with periodic boundary conditions.
This feature makes the SSD useful for 
more efficiently numerically finding the ground state in DMRG.

Similar to the rainbow deformation,
the M\"obius/SSD deformations can also be understood by
putting CFTs on a curved background with the metric
\cite{2018PhRvB..97r4309W}
\begin{align}
  ds^2_{{\it Mobius}}
  % &=
  % \left(
  %   \cosh 2\theta - \sinh 2\theta \cos \frac{2\pi x}{L}
  % \right)^2 
  % \frac{dt_E^2}{\cosh^2 2\theta}
  % +
  % dx^2
  % \nonumber \\
    =
      -\left(
      1 - \tanh 2\gamma \cos \frac{2\pi x}{L}
      \right)^2 
      dt^2
      +
      dx^2.
  \label{ds Mobius}
\end{align}

%\textcolor{red}{(Is there a name for this space?
%What is the penrose diagram?)}

\subsection{Overview of the paper}

In addition to the rainbow and M\"obius/SSD deformations,
various other examples
of inhomogeneous systems include
entanglement Hamiltonians,
\cite{hislop_longo_1982,
  2004JSMTE..12..005P,
  2016JSMTE..12.3103C}
the square root deformation
(known in the context of perfect state transfer),
\cite{2016PhRvB..93w5119W}
free fermions in harmonic traps and other potentials,
\cite{2016arXiv160604401D}
hyperbolic deformations,
and others.
\cite{
  new_calabrese_paper,
  2010arXiv1008.3458U,
  2010PThPh.124..389U,
  2010JPSJ...79j4001I,
  2018arXiv180710239L,
  2017arXiv170500679D}

In this work, we will discuss a
series of inhomogeneous (1+1)d systems,
which are given as deformations of uniform CFTs.
Of particular interest
is the scaling of entanglement entropy
at zero and finite temperatures,
which we obtain by
by combining
field theory,
holographic, and numerical approaches. 

As for the holographic approach, 
we develop holographic duals
of inhomogeneous (1+1)d CFTs
by finding appropriate foliations (slicings)
of the bulk ${\it AdS}_3$.
The simplest example would be to foliate
${\it AdS}_3$ by (1+1)d flat Minkowski spaces, 
which, at the asymptotic boundary,
gives rise to CFT on a flat space. 
Other foliations are also possible,
realizing CFT put on different metrics.  
For example,
${\it AdS}_3$ can be foliated by ${\it AdS}_2$,
\cite{2011JHEP...02..041A, 2014PhRvD..90f4019J, 2012JHEP...01..123A}
which, as we will discuss, realizes the rainbow chain
at the boundary.

%Then, it is natural to seek for the holographic description
%of a conformal field theory on $AdS_{2}$.
%The holographic dual of a CFT put on $AdS_{d+1}$ has been studied
%For example, see Ref.\ \onlinecite{2011JHEP...02..041A}.
%Also from a slightly different perspective in 
%Ref.\ \onlinecite{2014PhRvD..90f4019J} and Ref.\ \onlinecite{2012JHEP...01..123A}.

We will also discuss holographic duals of inhomogeneous CFTs at finite
temperatures,
by starting from the bulk spacetime with
the BTZ black hole, and following the same strategy as the
zero temperature case mentioned above.
We will show that sensible foliations, valid for low temperatures,
can easily be constructed by
a brute force application of the coordinate transformations
that we use to construct the foliations for the corresponding
zero-temperature geometry.
This naive approach, however,  
breaks down at higher temperatures,
and better foliations must be constructed to capture
the full temperature dependence of entanglement entropy.

%Our prescription to find holographic duals
%should apply for generic inhomogeneous CFTs,
%and may be applicable to 
%other interesting examples
%including entanglement Hamiltonians in CFTs,
%~\cite{hislop_longo_1982,
%  2004JSMTE..12..005P,
%  2008PhRvL.101a0504L,
%  2016JSMTE..12.3103C}
%the square root deformation
%which is related to perfect state transfer
%\cite{2016PhRvB..93w5119W},
%free fermions in harmonic traps and other potentials,
%\cite{2016arXiv160604401D}
%hyperbolic deformations,
%and others.
%\cite{new_calabrese_paper, 2010arXiv1008.3458U, 2010PThPh.124..389U, 2010JPSJ...79j4001I,2011PhRvL.107j1602T, 2018arXiv180710239L, 2018PTEP.2018f1B01T,
%2018JSMTE..04.3105T, 2018arXiv180500031W,
%2017PhRvB..96x5105M,
%2017arXiv170906238T,
%2017arXiv170500679D,
%2018PTEP.2018f1B01T,
%2018JSMTE..04.3105T,
%2018arXiv180500031W,
%2017PhRvB..96x5105M,
%2017arXiv170906238T,
%2017arXiv170500679D,
%2016JSMTE..12.3103C,
%2018PhRvB..97r4309W,
%2010arXiv1008.3458U,
%2010PThPh.124..389U,
%2010JPSJ...79j4001I,
%2010NJPh...12k3049V,
%2014JSMTE..10..004R,
%2015JSMTE..06..002R,
%2016arXiv161108559R,
%2011JHEP...02..041A,
%2014PhRvD..90f4019J}

In order to systematically derive 
suitable finite temperature geometries, 
we must solve Einstein's equations in the bulk
with a constant, negative cosmological constant,
with our curved $2d$ metric of interest given as a boundary condition.
Finding
solutions of the Einstein equation
with a prescribed boundary metric
is a well-studied problem.
See for example, Ref. \onlinecite{2000PhLB..472..316S}.
Our strategy as presented above
differs from the one in the aforementioned reference, 
but yields finite temperature results that agree very well with numerics.
We will however, remark briefly on the more systematic approach,
which involves solving Einstein's equations exactly for a given boundary metric.
The advantage of our foliation-based approach, is that we need not find new geodesics in our bulk geometry;
we can simply use the geodesics from the BTZ spacetime, being sure to implement an appropriate UV cutoff.

Finally, We will also look briefly at a larger class of deformations known as ``solution-generating diffeomorphisms".
In particular, we will discuss 
a holographic description of these inhomogeneous systems, and their entanglement properties. 

\section{Different foliations in AdS/CFT
and entanglement entropy}

In this section, we collect necessary ingredients to
describe the holographic duals
of rainbow and SSD deformations,
and the calculations of entanglement entropy.

\subsection{Different foliations in AdS/CFT}

In the AdS/CFT correspondence,
the CFT is defined (``lives'') on an asymptotic boundary of AdS.
What concerns us in developing our holographic description of
inhomogeneous systems is the fact that 
the AdS spacetime can be foliated (sliced)
in various different ways.
Different foliations (slicings)
give rise to different asymptotic boundaries,
and hence correspond to different situations on the CFT side. 
(See below, in particular Sec.\ \ref{Foliation and UV cutoff}.)
%\textcolor{red}{(Citations?)}

For example, consider the Poincar\'e patch of AdS
described by the metric:
\begin{align}
  \label{Poincare metric}
  ds^2_{{\it AdS}_3}
  &=
    R^2
    \frac{dz^2 + dx^2-dt^2}{z^2},
\end{align}
where
$z>0$ and $-\infty< t,x<+\infty$,
and $R$ is the radius of AdS.
(This coordinate patch covers only half of ${\it AdS}_3$.)
In these coordinates, ${\it AdS}_3$ is foliated by
$(1+1)$d flat spacetimes described by the coordinate $(t,x)$
(Fig.\ \ref{fig:rainbowContours}).
Realized at the asymptotic boundary $z\to 0$
is the CFT in its ground state
defined on the infinite line $-\infty < x < +\infty$.
Here, as usual, the central charge of the CFT is
$c= 3 R/(2 G_N)$ where $G_N$ is the Newton constant.

\subsubsection{Rindler evolution}

${\it AdS}_3$ has foliations other than the one suggested by the Poincar\'e metric.
Let us consider the Rindler-AdS foliation,
\cite{2012arXiv1211.7370P,
  2014arXiv1407.4467C,
  1998CQGra..15L..85D,
  2010JHEP...06..078C}
which can be obtained from the Poincar\'e
metric by the $z$-independent coordinate transformation,
\begin{align}
  \label{Rindler coord}
  t = u \sinh (h t'),
  \quad
  x = u  \cosh (h t'),
\end{align}
where $u>0$ and $-\infty < t'<\infty$.
This coordinate covers half of the original space, so that
the system is at a finite Unruh temperature.
The metric in this coordinate is 
\begin{align}
  \label{Rindler foliation}
  ds^2_{{\it AdS}_3} =
  \frac{R^2}{z^{2}}
  \left[
  (- u^2 dt^{\prime 2}+ du^{2})
  +
  dz^{2}
  \right].
\end{align}
It is also instructive to
introduce a tortoise coordinate
$x'\in [-\infty, +\infty]$
by
$u=: h^{-1}e^{ hx'}$.
The metric is
\begin{align}
  ds^2_{{\it AdS}_3} =
  \frac{R^2}{z^{2}}
  \left[
  e^{2 h x'} (-dt^{\prime 2}+ dx^{\prime 2})
  +
  dz^{2}
  \right].
\end{align}
Here, 
\begin{align}
ds^2_{{\it Rindler}} = e^{2 h x'} \left(-dt^{\prime 2}+ dx^{\prime 2}\right)
\end{align}
is the line element of the 2d Rindler space. 
%\textcolor{red}{(Cite papers)}
From Eq.\ \eqref{Rindler foliation},
we read off the Rindler Hamiltonian,
\begin{align}
  H_{{\it Rindler}} = \int^{\infty}_0du\, u\, \mathcal{H}(u).
\end{align}
This is nothing but the entanglement
Hamiltonian of the half space,
starting from the vacuum of CFT. 
(The entanglement Hamiltonian of the finite interval can be discussed
similarly.
\cite{2011JHEP...05..036C,2016JSMTE..12.3103C})

%*****
%
%If instead we use the global coordinate of AdS,
%where the metric is given by
%\begin{align}
%    ds^2_{{\it AdS}_3} =
%    - \left(  \frac{r^2}{L^2} +1\right)
%    dt^2
%    +
%    \left(  \frac{r^2}{L^2} +1\right)^{-1}dr^2
%    +
%    r^2 d\phi^2,
%\end{align}
%with $t\in[-\infty, \infty]$,
%$r>0$ and $\phi\in[0,2\pi]$,
%we realize,
%at $r\to \infty$,
%the ground state of the CFT put on
%a spacetime cylinder
%(the spatial part is a circle). 
%Examples of different foliations of AdS
%is presented in Table.\ **.
%(Considering $t\in[-\infty,+\infty]$,
%we have extended ${\it AdS}_3$ to its universal cover.)
%
%
%

In the Rindler foliation,
the coordinate system \eqref{Rindler coord}
covers only the half of the boundary.
In other words, observers in the coordinate patch \eqref{Rindler coord} 
cannot access the other half.
The observers are hence effectively at finite temperature.\cite{2011NuPhB.844....1H}
Correspondingly, in the bulk, there is a topological black hole.~\cite{2011JHEP...05..036C} Other foliations of $AdS_3$ lead to other inhomogeneous field theories on the boundary.

\subsubsection{Rainbow chain}

\begin{figure*}
    \centering
    \includegraphics[scale=0.32]{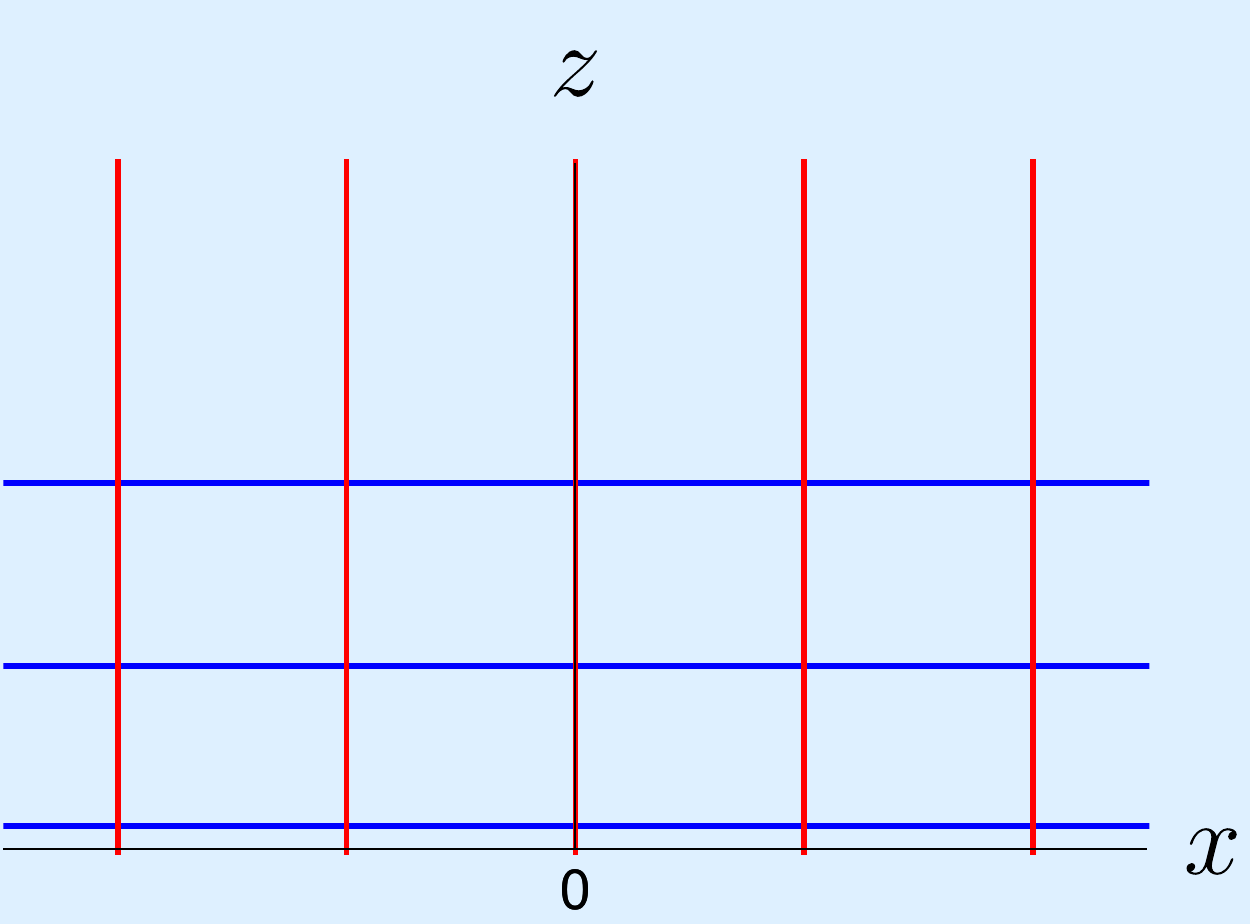}
    \hspace{0.5cm}
    \includegraphics[scale=0.32]{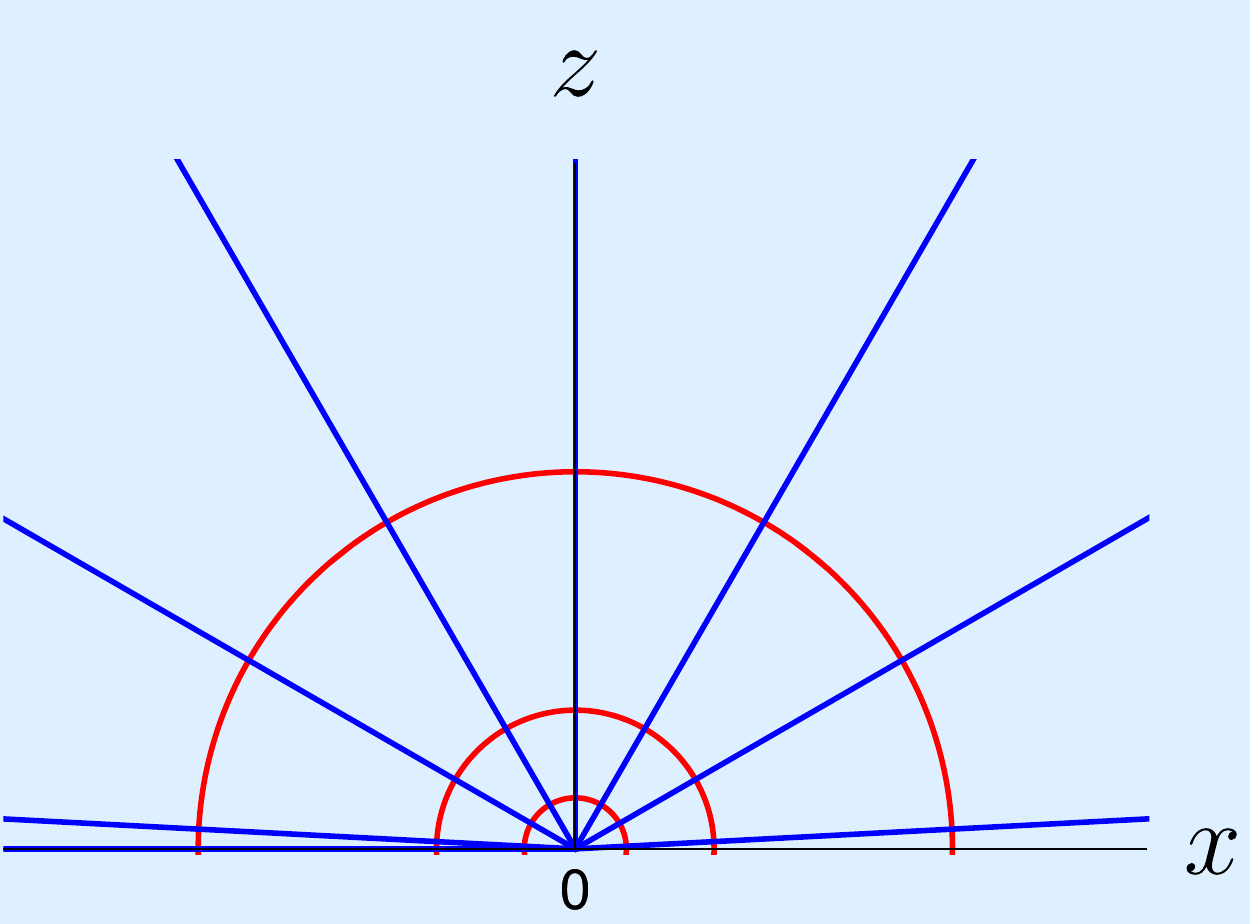}
    \hspace{0.5cm}
    \includegraphics[scale=0.32]{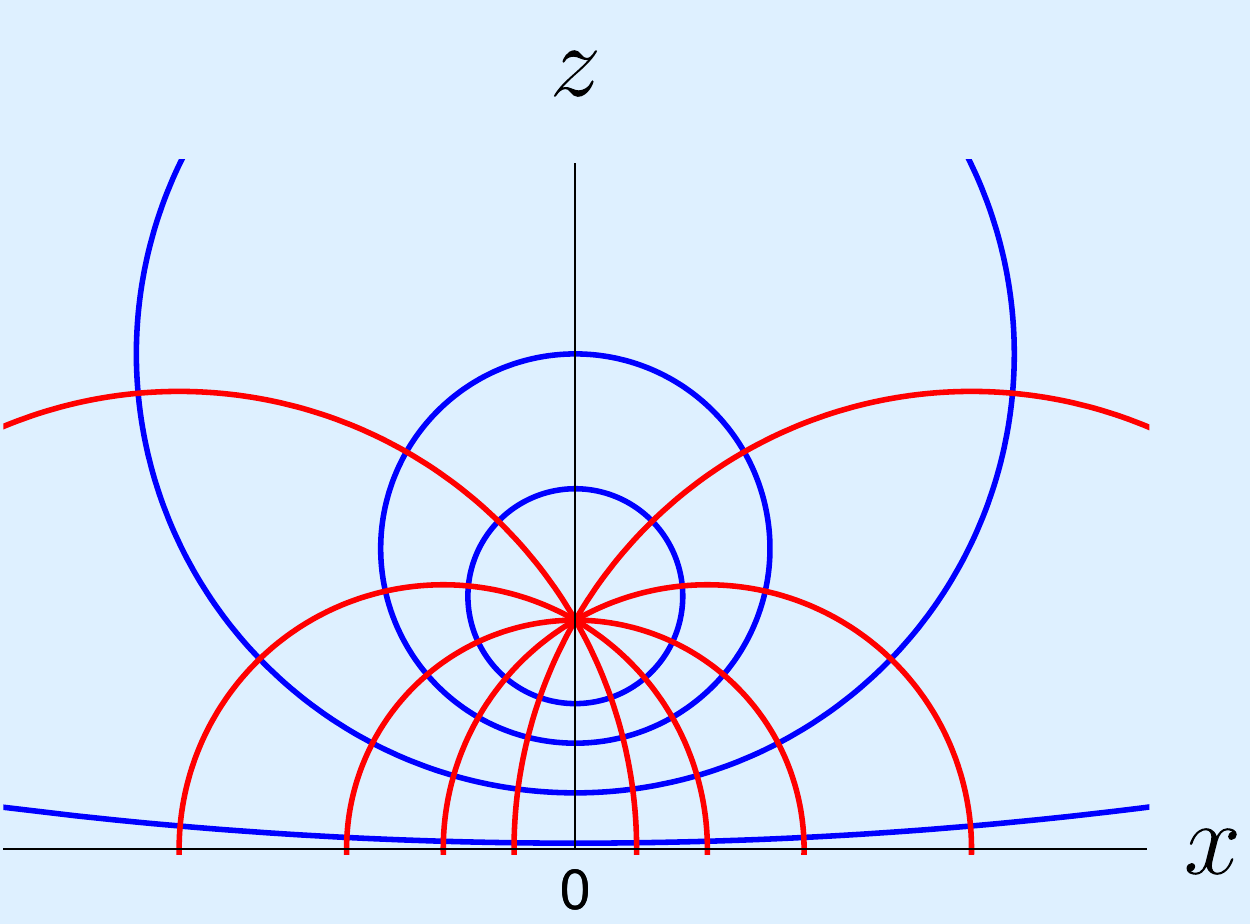}
    \caption{
      Three different foliations of ${\it AdS}_3$
      by flat Minkowski spaces (Left),
      ${\it AdS}_2$ (Middle),
      and 
      2d spaces with the metric \eqref{ds Mobius} (Left).
      (Middle)
      Lines of constant $\eta$ (red) and $\Theta$ (blue) for the rainbow
      coordinate transformation (\ref{rainCoord}); 
      (Right)
      Lines of constant $v$ (red) and $u$ (blue)
      for the SSD coordinate transformation (\ref{ssdCoord}) with $a=1$.
      Both are
      plotted in the original Poincar\'e spatial coordinates, $x$ and $z$. 
    }
      %A zero tension brane occurs at $\eta=0$, corresponding to the rainbow chain defect.}
    \label{fig:rainbowContours}
\end{figure*}

To realize the rainbow chain, 
we foliate ${\it AdS}_3$ by ${\it AdS}_2$.
\cite{2011JHEP...02..041A, 2014PhRvD..90f4019J, 2012JHEP...01..123A}
The corresponding metric can be obtained
from the Poincar\'e metric \eqref{Poincare metric}
by the following $t$-independent coordinate transformation:
\begin{align}
  z = \eta \cos (h \Theta),
  \quad
  x = \eta \sin (h \Theta),
  \label{rainCoord}
\end{align}
where $\eta>0$ and $-\pi/2h< \Theta <\pi/2h$. Contours of constant $\eta$ and $\Theta$ are plotted in Fig.\ \ref{fig:rainbowContours}.
The metric is given by
\begin{align}
  \label{AdS foliation}
  ds^2_{{\it AdS}_3} =
  \left[
  \frac{h^2 R^2}{\cos^2 (h \Theta)}
  \right]
  \left[
  d\Theta^2 + ds^2_{{\it AdS}_2}
%  \frac{1}{h^2 \eta^2}
%  \left(
%  d\eta^2 - dt^2
%  \right)
  \right],
\end{align}
where $ds^2_{{\it AdS}_2}$ is given by \eqref{ds AdS2}.
There are {\it two} asymptotic boundaries,
one at $\Theta=+\pi/2h$ and the other at $\Theta= -\pi/2h$.
%We recognize in \eqref{AdS foliation}
%the metric of ${\it AdS}_2$ with radius $1/h$,
%\begin{align}
%  ds^2_{{\it AdS}_2}
%  =
%  \frac{1}{h^2 \eta^2}
%  \left(
%  d\eta^2 - dt^2
%  \right).
%\end{align}
%Therefore,
There are two CFTs, one for each boundary,
which are put on ${\it AdS}_2$.
These two CFTs are not decoupled, but
are connected at the boundary of ${\it AdS}_2$.
The ground state is highly entangled between the two CFTs.

\subsubsection{M\"obius and SSD evolution}

%\begin{figure}
%    \centering
%    \includegraphics[scale=0.4]{SSDfoliation.pdf}
%    \caption{Lines of constant $v$ (red) and $u$ (blue) for the SSD coordinate transformation (\ref{ssdCoord}) with $a=1$ plotted in the original Poincar\'e spatial coordinates, $x$ and $z$. }
%    \label{fig:ssdContours}
%\end{figure}

The prescription of generating
the holographic dual of the
rainbow chain
%(= CFT on ${\it AdS}_2$)
can be generalized, by considering
different $t$-independent coordinate transformations
than (\ref{rainCoord}).
Let us now consider a coordinate map
%with two poles 
\begin{align}
u+iv= \log(z+ix+a)-\log(z+ix-a),
\end{align}
where $a$ is a real parameter.
Inverting this and separating the real and imaginary parts, we have:
\begin{align}
  z &= \frac{a\sinh u}{\cosh u-\cos v},
      \quad
  x = \frac{-a\sin v}{\cosh u - \cos v},
  \label{ssdCoord}
\end{align}
where $u \in [ 0,\infty)$ and $v \in [ 0,2\pi)$. This coordinate transformation is plotted in Fig.~\ref{fig:rainbowContours}. Our Poincar\'e metric thus becomes 
\begin{align}
  \label{AdS SSD}
  ds^2_{{\it AdS}_3}&=
        \frac{1}{\sinh^2 u}
        \left[ du^2 + dv^2
        - a^{-2} (\cosh u-\cos v)^2 dt^2 \right]
\end{align}
and our conformal boundary now occurs as $u \rightarrow 0$.
Near the boundary $u\to u_0$, the boundary metric is given by 
\begin{align}
  \label{SSD from AdS}
  ds^2&=
%        \frac{1}{\sinh^2 u_0}
%        \left[ 
%        - a^{-2} (\cosh u_0-\cos v)^2 dt^2
%        +dv^2
%        \right]
%        \nonumber \\
%  &=
        \left[
        \frac{1}{a \tanh u_0}
        \right]^2
        \left[
        - \left(1 - \frac{\cos v}{\cosh u_0}\right)^2 dt^2
        +\frac{a^2 dv^2}{\cosh^2 u_0}
        \right]
\end{align}
where $u_0$ is the UV cutoff.
To make contact with \eqref{SSD envelope},
we introduce a parameter $\gamma$ by
\begin{align}
  \tanh 2\gamma
  =
   \frac{1}{\cosh u_0}.
\end{align}
By further introducing $L$ by
\begin{align}
  \label{L vs a}
  \frac{L}{2\pi} =
  \frac{a}{\cosh u_0}
  \stackrel{u_0 \to 0}{\longrightarrow} a, 
\end{align}
and the change of variable
$v = 2\pi x/L$, we arrive at
\begin{align}
&  ds^2=
%        \left[
%        \frac{1}{a \tanh u_0}
%        \right]^2
%        \left[ 
%        -
%        \left(1-
%        \frac{1}{\cosh u_0}
%        \cos \frac{2\pi x}{L}
%        \right)^2 dt^2
%       + 
%        dx^2
%        \right]
%        \nonumber \\
%  &=
         \left[ \frac{\cosh 2\gamma}{a} \right]^2
  ds^2_{{\it Mobius}},
%        \left[ 
%        -
%        \left(1- \tanh 2\gamma 
%        \cos \frac{2\pi x}{L}
%        \right)^2 dt^2
%       + 
%        dx^2
%        \right].
%        \nonumber \\
\end{align}
where
$ds^2_{{\it Mobius}}$
is given by \eqref{ds Mobius}.
%  \nonumber \\
%  &
%  ds^2_{{\it Mobius}}
%  % &=
%  % \left(
%  %   \cosh 2\theta - \sinh 2\theta \cos \frac{2\pi x}{L}
%  % \right)^2 
%  % \frac{dt_E^2}{\cosh^2 2\theta}
%  % +
%  % dx^2
%  % \nonumber \\
%    =
%      -\left(
%      1 - \tanh 2\gamma \cos \frac{2\pi x}{L}
%      \right)^2 
%      dt^2
%      +
%      dx^2.
%\end{align}
%Here, the metric $ds^2_{{\it Mobius}}$
%is the one which corresponds
%to the M\"obius evolution.
%\cite{2018PhRvB..97r4309W}
%Here, $\gamma$ is a parameter;
%$\gamma=0$ corresponds to the evolution by
%the ordinary, uniform Hamiltonian,
%whereas $\gamma\to \infty$ corresponds to
%the SSD evolution. 
In our metric \eqref{SSD from AdS},
the UV cutoff $u_0$ plays the role of $\gamma$;
one can then see that our foliation
realizes a regularized version of the SSD.

%\textcolor{red}{(Q:
%  is there a coordinates system where
%  we can realize and tune $\gamma$ as a parameter,
%  rather than UV cutoff? Use transformation in appendix of 1802.07765
%  )}

\paragraph{Dipolar limit}
It is also interesting to take 
the limit $a\to 0$, while keeping $z/a$ and $x/a$ finite;
this is the dipolar limit.
We then consider the coordinate transformation:
\begin{align}
  u + iv = \frac{a}{z + ix}.
\end{align}
Separating into real and imaginary parts, we have, 
\begin{align}
x= \frac{-av}{u^2+v^2},
  \quad 
z= \frac{au}{u^2+v^2}.
\end{align}
The metric is then given by
\begin{align}
  \label{AdS SSD dipolar}
  ds^2_{{\it AdS}_3} = 
  \frac{R^2}{u^2}
  \left[
  dv^2
  - a^{-2}(u^2+v^2)^2dt^2
  + du^2  \right],
\end{align}
where $u\to 0$ corresponds to the conformal boundary.
Near the boundary $u\sim u_0$, 
the metric for a given slice is 
\begin{align}
  ds^2 =
  \frac{R^2}{u^2_0}
  \left[
  - a^{-2}(v^2+u^2_0)^2dt^2
  + dv^2  \right].
\end{align}

\subsubsection{Foliations and UV cutoff}
\label{Foliation and UV cutoff}

For each of the different coordinate transformations
we considered above,
we have an associated ``natural'' foliation; 
For example,
in \eqref{AdS foliation},
we have a family of surfaces with ${\it AdS}_2$ metric
parameterized by $\Theta$.
To properly define CFTs in the asymptotic boundaries, 
we further need to introduce a UV cut off.
We do so by taking our cutoff surface, i.e.,
the surface where we define our CFTs,
to be one of the slices located near (but away from) the
boundary (boundaries).
This is the UV cutoff which is ``consistent'' or ``natural''
for a given foliation.
In terms of the Poincar\'e coordinate \eqref{Poincare metric}
that we started with,
this means that our the cutoff is position-dependent
($x$-dependent).
Assuming our bulk foliation is dictated by a coordinate transformation,
$z=z(u,v)$ and $x=x(u,v)$ (where $u$ and $v$ are our new radial and transverse
coordinates, respectively), we replace the UV cutoffs
with their curvilinear counterparts:
\begin{align}
  \epsilon \to z(u=\epsilon,v(x))
  \simeq
  \frac{\partial z(u,v)}{\partial u} \bigg |_{\substack{u=0 \\ v=v(x)}}  \epsilon.
    \label{curvilinearCutoff}
\end{align}

It is worth emphasizing that
it is because of the cutoff that
we realize a ``different CFT''
as mentioned, e.g., in Ref. \onlinecite{2014arXiv1407.4467C}.
For example, 
the metric of the type \eqref{AdS foliation}
was previously used to discuss
holographic duals of boundary CFTs (BCFTs)
\cite{2011PhRvL.107j1602T,2011JHEP...11..043F}; 
In the ${\it AdS}/{\it BCFT}$ correspondence, 
one realizes a BCFT on ${\it AdS}_2$ which has a boundary
(or boundaries).
There, however, one imposes the ``original'' cutoff
using the flat Minkowski cutoff surfaces.\cite{2011PhRvL.107j1602T}

\subsubsection{Solution generating diffeomorphisms (SGDs)}

We can construct a family of locally ${\it AdS}_3$ spacetimes with the
appropriate asymptotic behavior (i.e. that preserve the form of the
Fefferman-Graham metric, up to gauge transformations) by acting on the vacuum
metric with a certain class of
diffeomorphisms.\cite{1999AIPC..484..147B,2010JHEP...06..078C} Applying one of
these transformations corresponds to exciting a state in the
CFT.\cite{2012JHEP...12..027R,2015JHEP...01..036M,2018arXiv180608376F} We say
therefore that they are ``solution-generating diffeomorphisms''.

Using light cone coordinates for the boundary, $x_\pm= t \pm x$, we can parameterize the transformations as follows:
\begin{align}
  x_{\pm}= f_{\pm}(\tilde{x}_{\pm}),
  \quad
%  \quad x_-= f_-(\tilde{x}_-), \quad
    z= \tilde{z} \sqrt{f'_+(\tilde{x}_+)f'_-(\tilde{x}_-)}.
\end{align}
Starting from the Poincar\'e metric \eqref{Poincare metric},
%\begin{align*}
%  ds^2_{{\it AdS}_3} = \frac{dz^2 - dx_+dx_-}{z^2},
%\end{align*}
we obtain the following bulk metric in the new coordinates:
\begin{align}
  ds^2_{{\it AdS}_3}
  &= \frac{d\tilde{z}^2 - d\tilde{x}_+d\tilde{x}_-}{\tilde{z}^2} + (A_+ d\tilde{x}_+ + A_- d\tilde{x}_-)^2
        \nonumber \\
      &\quad  + \frac{2 d \tilde{z}}{\tilde{z} }(A_+ d\tilde{x}_+ + A_- d\tilde{x}_-)
    \label{sgdMetric}
\end{align}
where
$A_\pm= - ({1}/{2}) {f''_\pm(\tilde{x}_\pm)}/{f'_\pm(\tilde{x}_\pm)}$. 

Although
these diffeomorphisms preserve the form of the metric and are therefore trivial
gauge transformations from the perspective of the bulk, they are nontrivial at
the asymptotic boundary. This nontriviality can be understood by observing that
the SGDs result in nonzero contributions to the boundary stress tensor from the Schwarzian derivative:
\begin{align}
  T_{\pm \pm}
  &= \frac{c}{48 \pi f'_{\pm}(\tilde{x}_{\pm})^2}
    \left[ 3 f''_{\pm}(\tilde{x}_{\pm})^2
    - 2f'_{\pm}(\tilde{x}_{\pm}) f'''_{\pm}(\tilde{x}_{\pm}) \right],
%          \nonumber \\
%  T_{--}&= \frac{c}{48 \pi f'_-(\tilde{x}_-)^2} \left( 3 f''_-(\tilde{x}_-)^2 - 2f'_-(\tilde{x}_-) f'''_-(\tilde{x}_-) \right),
          \nonumber \\
    T_{+-}&= 0.
\end{align}
With the stress-tensor in hand, we can write the metric (\ref{sgdMetric}) in Fefferman-Graham form:
\begin{align}
    ds^2_{{\it AdS}_3}&= \frac{dz^2}{z^2}  - \frac{1}{z^2} \Bigg [dx_+ dx_- \\& + \frac{z^2}{4} \left( L(x_+)dx_+^2 + \bar{L}(x_-)dx_-^2 \right)  \\&+ \frac{z^4}{16}  L(x_+)\bar{L}(x_-)dx_+ dx_-  \Bigg ], 
    \label{sgdFG}
\end{align}
where $L(x_+)= \frac{48 \pi}{c} T_{++}$ and $\bar{L}(x_-)= \frac{48 \pi}{c} T_{--}$ \cite{2000PhLB..472..316S}. The nonzero energy-momentum tensor means that we are in an excited state of the original CFT. Indeed, we can write this state explicitly by finding a unitary representation of our diffeomorphisms. For a diffeomorphism (written in complex coordinates $z=x+it$) $f(z)=\sum^{\infty}_{n=-\infty} \epsilon_n z^{-n+1}$, this state is given by $|\Omega \rangle_f= U(\epsilon)|0 \rangle$, where\cite{2018arXiv180608376F}
$
U(\epsilon)= \exp \left( \sum^{\infty}_{n=-\infty} \epsilon_n L_{-n} \right),
$
where, $L_{-n}$ are the standard Virasoro generators.

Note that $z$ and $x_\pm$ in (\ref{sgdFG}) are ${\it not}$ the same as the original Poincar\'e coordinates; they are chosen specifically to put the metric in Fefferman-Graham form. In this gauge, the metric has a horizon located at $z=z_H=2(L(x_+)\bar{L}(x_-))^{-1/4}$. As $z\to 0$, we see that we recover the flat Minkowski metric. To investigate our excited state, we must use a constant $\tilde{z}=\epsilon$ cutoff corresponding to the curvilinear cutoff
\begin{align}
    z= \epsilon \sqrt{f'_+(\tilde{x}_+)f'_-(\tilde{x}_-)}.
    \label{sgdCutoff}
\end{align}

%Using (\ref{sgdCutoff}), we can compute the entanglement entropy of an interval in the $\tilde{x}_\pm$ coordinates we obtain from performing the bulk SGD.\cite{2012JHEP...12..027R,2016PhRvD..94l6006S} The result is
%\begin{align}
%    S_A&(x_1,x_2) \nonumber \\
%    =& \frac{c}{12} \log \left[ \frac{\mathcal{L}(\tilde{x}_{1+},\tilde{x}_{2+})^2\mathcal{L}(\tilde{x}_{1-},\tilde{x}_{2-})^2}
%    {\epsilon^4 f'_+(\tilde{x}_{1+})f'_+(\tilde{x}_{2+})f'_-(\tilde{x}_{1-})f'_-(\tilde{x}_{2-})}\right],
%    \label{sgdEnt}
%\end{align}
%where $\mathcal{L}(\tilde{x}_{1\pm},\tilde{x}_{2\pm})$ is the proper length on the boundary between $\tilde{x}_{1\pm}$ and $\tilde{x}_{2\pm}$. 

The essential physics in both the SGD protocol and the prescription we have outlined in this paper is the same. In both cases, we pick a UV cutoff for our bulk spacetime that depends on the transverse Poincar\'e coordinates. In our prescription, we arrive at the curvilinear cutoff by performing a particular time-independent coordinate transformation, while in the SGD case, the cutoff emerges as a natural result of extending a conformal transformation into the bulk. A key difference between the two is that SGDs are in general time-dependent, since they are formulated in terms of light-cone coordinates. Indeed, the class of Weyl transformations that the SGDs induce on the UV cutoff surface is limited to those of the form $\exp\left(2 \phi (\tilde{x}_+,\tilde{x}_-) \right)= f'_+(\tilde{x}_+)f'_-(\tilde{x}_-)$, while in our prescription, the Weyl transformations depend purely on the transverse spatial coordinate (e.g. $\exp\left(2 \phi (u)\right)= 1/u^2$ for the rainbow chain). Nevertheless, SGDs can be used to construct many interesting foliations of ${\it AdS}_3$. Examples include the Rindler foliation mentioned previously, the Hopf
fibration\cite{2014JHEP...02..118A}, and various two-sided geometries mentioned
in
Ref.\ \onlinecite{2015JHEP...01..036M}.

\subsection{Entanglement entropy}

\subsubsection{Zero temperature}

%The method used to find (\ref{sgdEnt}) will be used throughout this paper.
%Namely,
Once we have obtained a foliation of ${\it AdS}_3$ corresponding to our
inhomogeneous system of interest,
we can use the Ryu-Takayanagi procedure to compute the bipartite entanglement
entropy of a particular interval on the boundary.
\cite{2006JHEP...08..045R,2006PhRvL..96r1602R,2007JHEP...07..062H}
We start by using the known result for the zero temperature holographic entanglement entropy for an interval $[x_1,x_2]$ on the asymptotic boundary of the Poincar\'e patch:
\begin{align}
    S_A^{holo}(x_1,x_2)= \frac{c}{3} \log \left[ \frac{x_2-x_1}{\sqrt{\epsilon_1}\sqrt{\epsilon_2}} \right].
    \label{sHolo}
\end{align}
As before, 
assuming our bulk foliation is dictated by a coordinate transformation, $z=z(u,v)$ and $x=x(u,v)$ (where $u$ and $v$ are our new radial and transverse coordinates, respectively), we replace the UV cutoffs, $\epsilon_1$ and $\epsilon_2$ with their curvilinear counterparts:
\begin{align}
    \epsilon_i \to z(u=\epsilon,v(x_i))= \frac{\partial z(u,v)}{\partial u} \bigg |_{\substack{u=0 \\ v_i=v(x_i)}}  \epsilon.
    \label{curvilinearCutoff}
\end{align}
Equation \eqref{sHolo} then becomes
\begin{align}
    S_A^{holo}(v_1,v_2)= \frac{c}{3} \log \left[ \frac{x(u=0,v_2)-x(u=0,v_1)}{\epsilon  \sqrt{\frac{\partial z(u=0,v_1)}{\partial u} \frac{\partial z(u=0,v_2)}{\partial u}}} \right].
\end{align}

For example, in the dipole foliation, where $z=\frac{au}{u^2 + v^2}$,
the UV-cutoff and transverse coordinate transform as
\begin{align}
    \epsilon_i= \frac{a}{v_i^2}\epsilon, \quad x_i=x(u=0,v_i)=\frac{-a}{v_i}.
\end{align}
We can plug these into (\ref{sHolo}) to find the holographic entropy of the
dipole foliation.
%Overall, we will see that there is excellent agreement between the zero temperature holographic results and the zero temperature free fermion numerical results.
Similarly,
for metric \eqref{sgdMetric},
using the cutoff \eqref{sgdCutoff},
we can compute the entanglement entropy of an interval
in the $\tilde{x}_\pm$ coordinates:
\cite{2012JHEP...12..027R,2016PhRvD..94l6006S}
%we obtain from performing the bulk SGD.
%The result is
\begin{align}
    S_A&(x_1,x_2) \nonumber \\
    =& \frac{c}{12} \log \left[ \frac{\mathcal{L}(\tilde{x}_{1+},\tilde{x}_{2+})^2\mathcal{L}(\tilde{x}_{1-},\tilde{x}_{2-})^2}
    {\epsilon^4 f'_+(\tilde{x}_{1+})f'_+(\tilde{x}_{2+})f'_-(\tilde{x}_{1-})f'_-(\tilde{x}_{2-})}\right],
    \label{sgdEnt}
\end{align}
where $\mathcal{L}(\tilde{x}_{1\pm},\tilde{x}_{2\pm})$ is the proper length on the boundary between $\tilde{x}_{1\pm}$ and $\tilde{x}_{2\pm}$.

\subsubsection{Finite temperature}

Computing holographic entanglement entropy for inhomogeneous systems at finite
temperature is a less trivial matter than the zero temperature case;
we need to find a proper foliation of the BTZ black hole solution 
with a given metric on the boundary.
While this procedure can be implemented
in certain cases (as described below),
we will use a different approach in the bulk of the paper,
in which we cut off the BTZ black hole spacetime
%\cite{2000PhR...323..183A,1992PhRvL..69.1849B}
with the same curvilinear UV cutoff used in the previous section.
Although approximate,
this approach allows us to use known results for geodesic lengths of entangling
surfaces in the BTZ spacetime.
This will allow us to avoid solving for potentially complicated geodesics
that the bulk metrics from
the exact treatment would yield.

%
%Two possible approaches to the finite temperature problem are ($i$) to find the most general possible solution to Einstein's equations with constant negative curvature with a given static metric on the boundary, and ($ii$) to cut off the BTZ black brane spacetime\cite{2000PhR...323..183A,1992PhRvL..69.1849B} with the same curvilinear UV cutoff used in the previous section. We will use approach ($ii$) in this paper, but here we also outline approach ($i$) for completeness. Although the first approach is more systematic, the second allows us to use known results for geodesic lengths of entangling surfaces in the BTZ spacetime. This will allow us to avoid solving for potentially very complicated geodesics that the bulk metrics from ($i$) would yield.

Let us first outline the exact approach
following the results in Ref.\ \onlinecite{2000PhLB..472..316S}.
We begin by specifying an arbitrary static (1+1)d boundary metric on the $x-t$ plane:
\begin{align}
    ds^2_\partial= -h(x) dt^2 + \frac{dx^2}{h(x)}.
    \label{arbBdyMet}
\end{align}
Note that any (1+1)d static metric can be written this way, up to a spatial
coordinate transformation.
We can solve the resulting Dirichlet problem in the Fefferman-Graham gauge in terms of two undetermined constants, $J$ and $B$. Assuming $J=0$ (to avoid a cross term in the metric), the resulting bulk metric is
\begin{align}
  ds^2 &= \frac{R^2 dz ^2}{z ^2}
         + \frac{1}{z^2} \bigg[
         -h\left( 1+ \frac{R^2}{16}\frac{h'^2-B^2}{h}z^2 \right)^2 dt^2
         \nonumber \\
       &\quad
         +
         \frac{1}{h}  \left( 1 + \frac{R^2}{4} h'' z^2 - \frac{R^2}{16}\frac{h'^2-B^2}{h}z^2  \right)^2  dx^2
         \bigg].
    \label{dirichletSoln}
\end{align}
For $h(x)=1$, the above reduces to the Fefferman-Graham form of the BTZ
metric\cite{2012JHEP...02..054T} with a horizon at
$z_H= {4}/{R B}$. The parameter $B$ thus corresponds to temperature in the
boundary CFT.

For a more general $h(x)$, a more interesting horizon will be present. For
example, if we start with the ${\it AdS}_2$ boundary metric
 \eqref{ds AdS2}
%we expect to see in the rainbow chain (suppressing the curvature scale),
%\begin{align}
%    ds^2_\partial= \frac{d\eta^2 - dt^2}{\eta^2},
%\end{align}
making the coordinate change $x= 1/\eta$ puts the metric in the form of
(\ref{arbBdyMet}) with $h(x)= x^2$
(here we set the curvature scale $h$
in \eqref{ds AdS2} to be 1 for simplicity).
Plugging this into (\ref{dirichletSoln}) and transforming back to the original $\eta$ coordinate gives us the following bulk solution:
\begin{align}
    ds^2= \frac{dz^2}{z^2} + \frac{1}{ z^2 \eta^2} \bigg [ \left( 1+R^2 (1+B^2\eta^2)z^2 \right)^2 dx^2 \nonumber \\  
    - \left( 1+R^2 (1-B^2\eta^2)z^2 \right)^2 dt^2 \bigg ],
    \label{rainbowBulk}
\end{align}
which appears to have an interesting horizon at
$z= {1}/({R\sqrt{B^2\eta^2-1}})$. Note that as $z\to 0$, we see our bulk metric
reduce to the ${\it AdS}_2$ foliation that we expect from the rainbow chain
dual.

Computing the entanglement entropy would entail picking two boundary points,
$\eta_1$ and $\eta_2$, on a fixed time slice of (\ref{rainbowBulk}), and
computing the length of the spacelike geodesic between them. This would of
course require a UV cutoff, which we would choose to be a constant $z=\epsilon$.
Solving the Euler-Lagrange equations for the spacelike geodesics of
(\ref{rainbowBulk}) is nontrivial, and will be different for each inhomogeneous
system of interest.
It is for this reason that we use
the approximate approach in this paper.

This approximation for finding the
finite temperature entanglement entropy in inhomogeneous systems requires two inputs. The first is the well-known result for the finite temperature holographic entanglement entropy\cite{2007JHEP...07..062H}, which in Poincar\'e coordinates takes the form
\begin{align}
    S^{holo}_A (x_1,x_2;\beta)= \frac{c}{3} \log
      \left[
      \frac{ \beta}{\pi \sqrt{\epsilon_1} \sqrt{\epsilon_2}}
      \sinh\left(\frac{\pi (x_2-x_1)}{\beta} \right) 
      \right].
    \label{BTZgeoLen}
\end{align}
For reference, in Poincar\'e coordinates the BTZ black hole metric is
\begin{align}
  ds^2_{{\it BTZ} }
  &=
  R^2 \left[
  -\frac{f(z)dt^2}{z^2}
  +
  \frac{dz^2}{f(z)z^2}
  +
  \frac{dx^2}{z^2}
    \right],
    \nonumber \\
  f(z) &= 1 - \frac{z^2}{z^2_H},
\end{align}
where $z_H$ is related to the inverse temperature $\beta$
at the boundary via $\beta = 2\pi z_H$.

With (\ref{BTZgeoLen}) in hand, we can add the second ingredient, and replace $\epsilon_1$ and $\epsilon_2$ with appropriate curvilinear cutoffs. As in the previous subsection, we can find these cutoffs from the bulk coordinate change that produced our zero temperature foliation. Combining (\ref{curvilinearCutoff}) and the coordinate transformations on the boundary with (\ref{BTZgeoLen}) yields a result for the finite temperature bipartite entanglement entropy in an inhomogeneous CFT:
\begin{align}
    &S^{holo}_A (v_1,v_2;\beta) \nonumber \\ &= \frac{c}{3} \log
      \left[\frac{\beta}{\pi \epsilon}
      \frac{ \sinh\left(\frac{\pi}{\beta} (x(u=0,v_2)-x(u=0,v_1)) \right) }{ \sqrt{\frac{\partial z(u=0,v_1)}{\partial u} \frac{\partial z(u=0,v_2)}{\partial u}}}
      \right].
      \label{finiteTempMaster}
\end{align}
The above can be considered our finite temperature master formula for entanglement entropy, and we will use it throughout the paper.

Though this approach is quite simple, it has one primary limitation. The
foliations that we use in vacuum ${\it AdS}_3$ are not always well-suited for
the black hole spacetime. Indeed, the curved cutoffs we use in the BTZ spacetime
will often collide with the black hole horizon. For example, the dipole
foliation uses a cutoff at $\epsilon_z=a\epsilon/v^2$. For a BTZ spacetime with
a horizon at $z=z_H$, this cutoff surface will actually be inside of the black
hole horizon for $v < \sqrt{a\epsilon/z_H}$. This means that a large part of our
boundary system could be inaccessible at very high temperatures (very small
$z_H$). This issue will become apparent in the ensuing comparisons between
numerics and holographic results, which will diverge from each other for
sufficiently large temperatures and in for intervals in certain regions of the
boundary. However, for most of the boundary theory, and for a wide range of
temperatures,
the approximation appears to yield results that agree quite well with numerics. We will thus use it in the remainder of this paper.

\iffalse
Connecting our results to these known results should help us find a finite temperature description of the rainbow chain and other inhomogeneous systems.

One can find a holographic description of the finite temperature BCFT physics by examining the behavior of the defect brane in thermal $AdS$ and BTZ spacetimes.~\cite{2011PhRvL.107j1602T} For a zero tension brane, the Hawking-Page transition between the two spacetimes occurs as it would without the brane. This justifies our use of the BTZ spacetime to describe the phenomenology of the rainbow chain at sufficiently high temperatures.

Finite temperature geometries
in the presence of defects have been also discussed,
which can be applied to discuss the properties of
the rainbow chain at finite temperature.
We start from the BTZ black hole (valid for higher temperature),
\begin{align}
  ds^2_{{\it AdS}_3}
  &=
  R^2 \left[
  -\frac{f(z)dt^2}{z^2}
  +
  \frac{dz^2}{f(z)z^2}
  +
  \frac{dx^2}{z^2}
    \right],
    \nonumber \\
  f(z) &= 1 - (z/z_H)^2,
\end{align}
where $z_H$ is related to the inverse temperature $\beta$
at the boundary as $\beta = 2\pi z_H$.
\fi

\subsection{CFT on curved spacetime}

The connection between the entanglement scaling
of inhomogeneous and homogeneous CFTs
can be also seen without using holography.
Recall that the entanglement entropy for a single interval
is given by the 2-point function of twist operators
located at two points $u_1$ and $u_2$ at a fixed time.
Let us consider the correlators
of arbitrary operators, $O_i(x_i)$, of CFT
put on the flat Euclidean metric
$ds_0(x)^2= dx^2 + d\tau^2$,
and the correlators of the same CFT 
put on the curved metric
\begin{align}
  ds^2
  &= e^{2\sigma(x)}ds_0^2
  = e^{2\sigma(x)} d\bar{z}dz
    \nonumber \\
  &=
    e^{2\sigma(x(u))}
  \left[  \left(  \frac{dx}{du} \right)^2  du^2 + d\tau^2\right], 
\end{align}
where $z=x+i\tau$ and $u$ is a spatial coordinate appropriate for the curved metric.
CFT gives us the following relationship between
correlation functions under Weyl and coordinate transformations:
\begin{align}
  &\langle O_1(u_1) O_2(u_2) \cdots \rangle_{ds^2}
  \nonumber \\
  &= e^{-\sum_i \Delta_i \sigma (u_i(x_i))} \langle O_1(x_1) O_2(x_2)
    \cdots \rangle_{ds_0(x)^2}
    \nonumber \\
  &= e^{-\sum_i \Delta_i \sigma (u_i)}
    \langle O_1(u_1) O_2(u_2) \cdots \rangle_{ds_0(u)^2},
    \label{cftCurved}
\end{align}
where $\Delta_i$ are the scaling dimensions of the operators,
which for the $n$-fold twist operators are 
\begin{align}
\Delta^{(n)}= \frac{c}{12} \left( n - \frac{1}{n} \right).
\end{align}
Since $ds_0(x)^2$ is just the Euclidean metric,
we can write down the known result for the two-point function of the twist operators at a fixed time (assuming we are on the infinite complex plane):
\begin{align}
  \label{twist operator in flat space}
  \mathrm{Tr}\,(\rho^n_A)\big |_{ds^2_0}
  &= 
\left\langle
    \sigma_n(x_1,\tau=0) \sigma_{-n}(x_2,\tau=0)
\right\rangle_{ds^2_0}
    \nonumber \\
  &= c_n
    \left[
    \frac{x_1-x_2}{\epsilon} \right]^{-2\Delta^{(n)}},
\end{align}
where $c_n$ is a constant coefficient that will make an $\mathcal{O}(1)$
contribution to the entanglement entropy.
The (Renyi) entanglement entropy in the curved background follows from
this expression by multiplying the Weyl factor
and transforming back to the original coordinates.

%\begin{center}
%\textcolor{red}{(finite T?)}
%\end{center}

\section{Holographic dual of rainbow chain}

We now discuss the holographic dual of the rainbow chain,
\eqref{AdS foliation},
in some detail. 
In particular, we study the entanglement entropy
for a given connected region $A$ of the boundary.  
As the system is inhomogeneous,
not only the size of the subregion,
but also its location matters.  
Here, we mainly consider two situations:
\begin{itemize}
\item
  ``Defect entanglement'':
  where we consider an entangling cut
  that separates the origin (``defect'') from the rest of the system.  

   \item
    ``Half chain entanglement'':
    where we consider an entangling cut
    that emanates from the origin (defect).
    This entangling cut thus splits
    the original system into two halves.
\end{itemize}
In addition, both of these two situations can be
studied at finite temperature. 

%By the holographic formula of entanglement entropy,
%the defect and half-chain entanglement entropies
%can be computed from the length of the corresponding geodesics.
%Recall that our holographic duals for inhomogeneous
%systems are obtained simply by taking
%appropriate foliations,
%i.e., by making appropriate coordinate transformations. 
%Thus, by construction, 
%relevant geodesics in our holographic duals
%are obtained
%from corresponding geodesics given in the Poincare coordinates
%by the coordinate transformations.
%One would then conclude that the scaling of the entanglement
%entropy as a function of the size of subsystem $A$
%would not be any different from the homogeneous case. 
%This naive conclusion however is not correct,
%because of the UV cutoff.
%We will impose the UV cutoff, such that it is
%consistent with a given foliation.
%As the system is inhomogeneous,
%the cutoff is position-dependent. Because of the cutoff,
%we realize a ``different CFT''
%as mentioned, e.g., in Ref.\ \onlinecite{2014arXiv1407.4467C}.

%  This geometry is appropriate
%  if one wishes to impose the so-called
%  transparent boundary conditions. 

%\begin{figure}
%  \includegraphics[scale=0.5]{slicing.png}
%\end{figure}

\subsection{Defect entanglement at zero temperature}

Let us first discuss the defect entanglement. 
%We expect that the defect entanglement is constant (independent of the size of the subsystem).
Recall the metric \eqref{AdS foliation} for the {\it AdS} foliation.
%\begin{align}
%  ds^2_{{\it AdS}_3} =
%  \left[
%  \frac{h^2 R^2}{\cos^2 (h \Theta)}
%  \right]
%  \left[
%  d\Theta^2 +
%  \frac{1}{h^2 \eta^2}
%  \left(
%  d\eta^2 - dt^2
%  \right)
%  \right].
%\end{align}
We choose, in the Poincare coordinates,
$x\in [-\eta_0, +\eta_0]$ on the boundary as the region of our interest.
%Focusing on the half of the ${\it AdS}_3$ with $\cos\Xi >0$, 
The geodesic $\Gamma$  anchored at $(\eta,\Theta)=(\eta_0, \pm \pi/2 h)$
is a semi-circle on the $z$-$x$ plane, i.e., $\eta(\Theta)={\it const.}=\eta_0$.
%Parameterizing $\Gamma$ by $\eta(\Xi)$, 
%The geodesic $\Gamma$ can be parameterized as $(\Xi(s), \eta(s))$. 
%We can take $\Xi$ as a parameter along $\Gamma$, and consider $(\Xi, \eta(\Xi))$.
%We seek the RT surface for the interval $0 < \eta < \eta_0$ on the boundary of $AdS_3$.
%The length of $\Gamma$ is given by
%
The length of $\Gamma$, $Len(\Gamma) = \int_{\Gamma} ds$, is given by
\begin{align}
  Len(\Gamma)
% % &= \int_{\Gamma} ds
% %  \nonumber \\
%   &=
%   \int \sqrt{ \frac{h^2R^2}{\cos^2 (h\Theta)}
%   \left( d\Theta^2 + \frac{ d\eta^2 }{h^2 \eta^2} \right)
%   }
%     \nonumber \\
%     &=
%     \int
%     \frac{R}{\cos \Xi}
%     \sqrt{
%     [d\Xi^2 + \frac{h^2 d\eta^2 }{\eta^2} ]
%   }
%     \nonumber \\
     &=
       hR
       \int^{\pi/2h}_{-\pi/ 2h}
     \frac{ d\Theta}{\cos (h\Theta)}
     \sqrt{
     1 + \frac{1 }{h^2 \eta^2} 
     \left(
       \frac{d\eta}{d\Theta} \right)^2
   }.
\end{align}
This integral is divergent, and one has to 
introduce a cutoff,
$\pm \pi/2h \to \pm (\pi/2h - \epsilon)$,
where $\epsilon$ is an $\eta$-independent constant.
The regularized length is a constant independent of $\eta_0$ 
(i.e., independent of the size of the subsystem):
\begin{align}
    S_A(\eta_0)= \frac{c}{3}\log\left( \frac{2}{\epsilon h} \right) + \mathcal{O}(1)
    \label{zeroTempDefect},
\end{align}
This agrees with the known
behavior of the defect entanglement entropy in the rainbow chain.
%We see that the defect entanglement is constant at zero temperature, as expected for the rainbow chain.
It does, however, depend on $1/h$, the length scale introduced by the rainbow defect (i.e., the $AdS_2$ radius). One can check that this is the same result we would have obtained using (\ref{sHolo}).

The ``defect entanglement" (\ref{zeroTempDefect}) we have just encountered resembles the constant ``boundary entanglement" found in a BCFT. The latter, however, is an $\mathcal{O}(1)$ constant correction to the logarithmically divergent leading order term, whereas the defect entanglement is constant, but $\mathcal{O}(\log(\epsilon))$. Indeed, in ${\it AdS/BCFT}$, a zero tension brane anchored at a boundary point of ${\it AdS}_3$ is dual to a half-space BCFT on the conformal boundary of the spacetime.~\cite{2011JHEP...11..043F} This particular ${\it AdS/BCFT}$ setup strongly resembles our effective holographic description of the rainbow chain, and indeed, ${\it AdS}_2$ foliation of the bulk captures the breakdown of the global symmetry group from $SO(2,2)$ to $SO(2,1)$. However, a zero tension brane (which is just an artifact of a coordinate transformation) yields zero boundary entropy, and the defect entanglement simply reflects the contributions of the ends of our entanglement interval to the entropy. A boundary entropy would reflect the entropy due to the presence of a tensionful brane. Adding such a brane could prove to be an interesting extension of the rainbow chain.

\subsection{Defect entanglement at finite temperature}

\begin{figure}[t]
\includegraphics[scale=0.2]{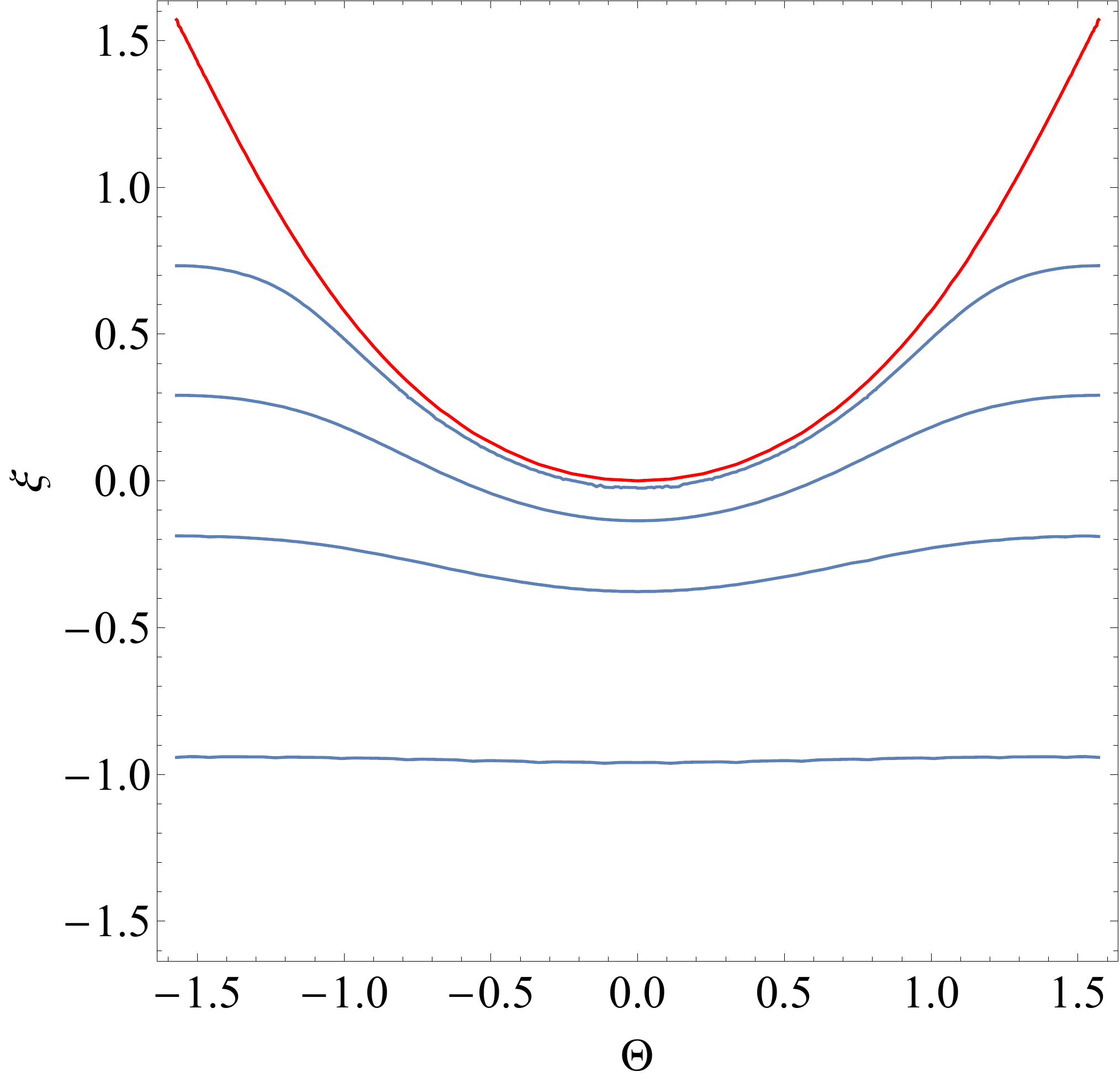}
\caption{
  \label{Horizon and geodesics}
  The black hole horizon (red) and several defect-crossing geodesics (blue)
  plotted in $\xi$ and $\Theta$ coordinates
  where
%  $\eta = \tan \xi + \sec \xi$
  $d\xi/\cos \xi = d\eta/\eta$
  (see Eq.\ \eqref{theta and xi}).
}
\end{figure}

We can bring this metric back to the one we
want to work with,
Eq.\ \eqref{rainCoord}.
%\begin{align}
%  z = \eta \cos (h\Theta),
%  \quad
%  x = \eta \sin (h\Theta),
%  \quad
%  t=t
%\end{align}
%where $\eta \in (0,\infty)$
%and the conformal boundaries are located at
%$\Theta=\pm \frac{\pi}{2h}$.
For reference, the geodesic in these coordinates takes the form
\begin{align}
  \sqrt{1-\frac{\eta^2 \cos ^2(h\Theta)}{z_H^2}}
  =
  \sqrt{1-\frac{z_*^2}{z_H^2}}
  \cosh \left[\frac{\eta \sin (h \Theta)}{z_H} \right].
\end{align}
In Fig.~\ref{Horizon and geodesics},
we have plotted the geodesic for several values of $z_*$
in blue, with the black hole horizon plotted in red.

%The length of the geodesic anchored at $\pm x_0$ on the Poincar\'e boundary is\cite{2007JHEP...07..062H}
%\begin{align}
%  l_{RT}= 2 R \log
%  \left[
%  \frac{2 z_H}{\epsilon_z} \sinh
%  \left(\frac{ x_0}{z_H} \right)
%  \right],
%  \label{finiteTLen}
%\end{align}
%where the spatial slice is cut off at $z=\epsilon_z$.
%\textcolor{red}{At first glance, it appears that even at low temperature (large $z_H$), the above expression will depend on $x_0$, which would disagree with the constant entanglement entropy seen in intervals symmetric about the defect in the rainbow chain. However, if we change our cutoff scheme to reflect our change of coordinates, we will see that the above geodesic length remains constant in the low temperature limit. 
%}

We will cut our spatial slice off at a constant
$\Theta_\epsilon = {\pi}/{2 h}-\epsilon $.
Expressed as a cutoff in the $z$ coordinate,
the cutoff now depends on $x_0=\eta_0$ as follows:
\begin{align}
  \epsilon_z=\eta_0 \cos (h \Theta_\epsilon )
  \approx \eta_0 h \epsilon.
\end{align}
Thus, using our master formula, (\ref{finiteTempMaster}), our defect entanglement entropy reduces to
\begin{align}
  S_A(\eta_0;\beta) &=
%      \frac{l_{RT}}{4 G_N} =
      \frac{c}{3} \log
      \left[
      \frac{ \beta}{\pi \epsilon h \eta_0}
      \sinh\left(\frac{ 2\pi \eta_0}{\beta} \right) 
      \right].
      \label{defectFinite}
\end{align}
Taking the zero temperature limit, $\beta \to \infty$,
we recover \eqref{zeroTempDefect}.
%we have
%\begin{align}
%  S_A(\eta_0)
% % &\approx \frac{c}{3} \log
% %     \left[
% %     \frac{ \beta }{ \pi\epsilon  h\eta_0}
% %       \left( \frac{2\pi \eta_0}{\beta } + \mathcal{O}(\beta^{-3})
% %       \right)
% %     \right]
%      \nonumber \\
%      &\approx 
%\frac{c}{3} \log \left(\frac{2}{\epsilon h} \right),
%\end{align}
%and we see that the defect entanglement is constant at zero temperature, as expected for the rainbow chain. It does, however, depend on $1/h$, the length scale introduced by the rainbow defect (i.e., the $AdS_2$ radius).
The proper length of the boundary interval
\begin{align}
    \ell = \frac{2}{h} \int^{\eta_0}_\epsilon \frac{d \eta}{\eta}.
\end{align}
In terms of the proper length, the defect entanglement entropy is
:
\begin{align}
S_A (x; \beta, \epsilon)
  = \frac{c}{3}
  \log \left[ \frac{\beta}{\pi h e^{h \ell}}
  \sinh \left( \frac{2 \pi \epsilon e^{h\ell}}{\beta } \right) \right]
+\cdots,
      \label{defectFinite properlength}
\end{align}
where
$\cdots$ is a non-universal part depending on the UV cutoff.

\begin{figure}[t]
  \includegraphics[scale=0.45]{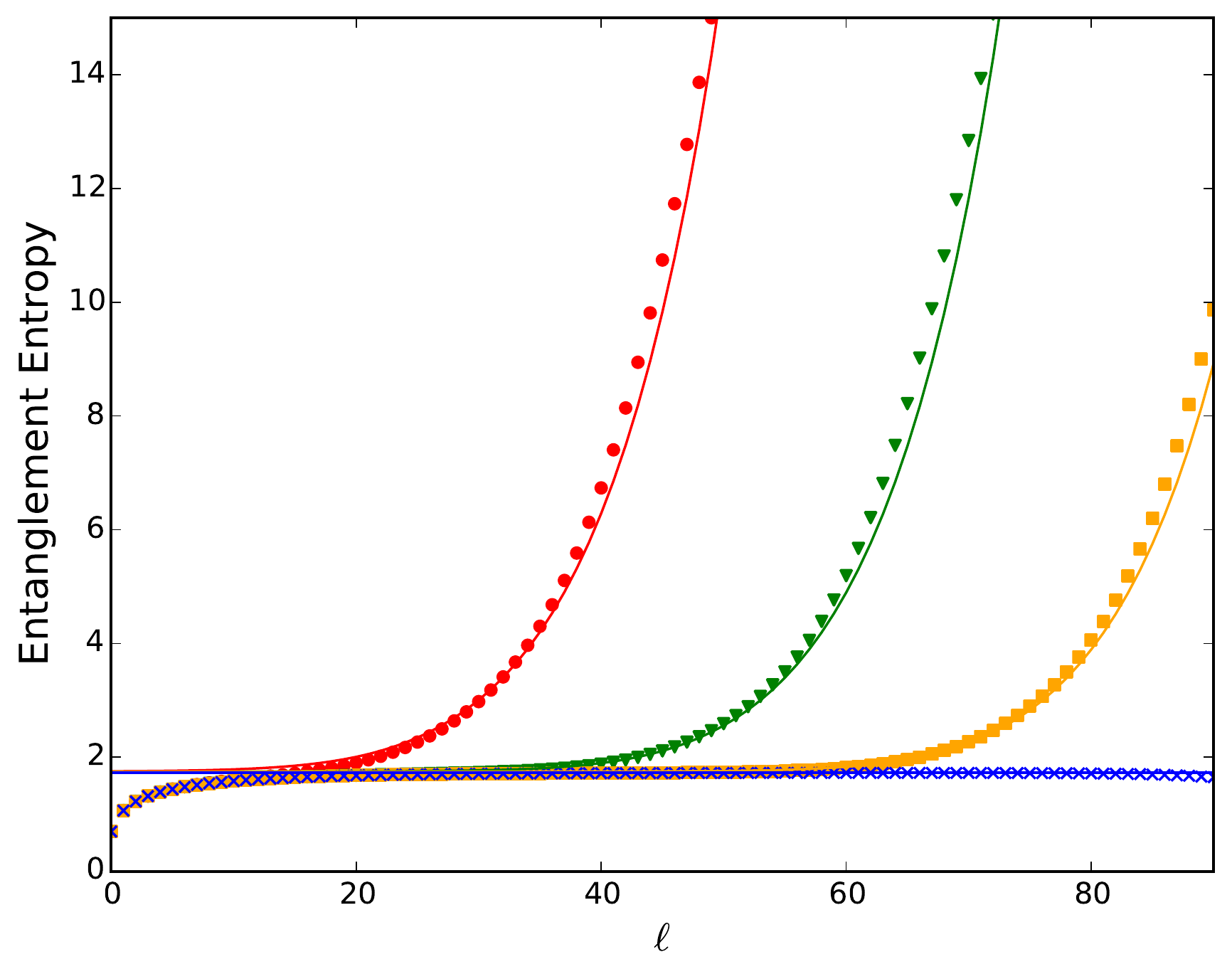}
  \caption{
    The numerically computed ``defect'' entanglement entropy
    %at $T=0$
    %for $h=0.5, 0.1, 0.05, 0.01$ from the top. 
    as a function of the proper length of the interval about the defect
    at $h=0.1$ and $\epsilon=0.5$
    and at finite temperatures
    $\beta=100, 1000, 10000, \infty$ from the top.
    The numerical data are fitted to \eqref{defectFinite properlength},
    where we treat the non-universal constant part
    (independent of $\ell$ and $\beta$)
    as a fitting parameter.
%    $S_A (x; \beta,b)
%    = \frac{c}{3}\left( \log \left[ \frac{\beta}{\pi e^{h |x|}} \sinh \left( \frac{2 \pi e^{h|x|}}{\beta h} \right) \right] + b \right),$
%where $b=2.25$ appears to give the best fit. The fit between the analytic an numerical results here is rather poor, revealing the limitations of our holographic scheme.
%\textcolor{red}{(x axis is not really the subsystem size.)}
  }
  \label{fig:defectPlot}
\end{figure}

\paragraph{Comparison with numerics}
To verify the finite temperature entanglement results for the rainbow chain, we
can numerically compute the entanglement entropy of an interval of space at a
given temperature for
the Hamiltonian \eqref{numerH}.
%the following Hamiltonian:
%\begin{align}
%    \hat{H}= \sum_{i,j=1}^N T_{ij} \hat{c}^\dagger_i \hat{c}_j,
%\quad 
%    T_{ij}= -f_i\delta_{i-j,1} -f_i \delta_{i-j,- 1}
%    \label{numerH}
%\end{align}
%where
%\begin{align}
%f_j=  \exp \left(  -h  |j -N/2|  \right)
%\end{align}
%in the case of a rainbow chain centered about $N/2$, and the $c_i$ operators are spinless fermions. 
%This is a lattice version of the $c=1$ free fermion CFT on a curved background with the
%metric $ds^2= -\exp \left( - 2h |x|  \right) dt^2 + dx^2$ (where the center has been shifted to $x=0$).
All numerical computations were done with open boundary conditions.
In Fig.~\ref{fig:defectPlot} we have plotted the numerically-computed
entanglement entropy for an interval symmetric about the defect at the origin of
the chain.
We can confirm the rapid asymptote at zero temperature of this
entropy to a constant, independent of the length of the interval, but dependent
upon $h$, the rainbow curvature scale.
Also plotted in Fig.\ \ref{fig:defectPlot}
is the defect entanglement for several different temperatures,
fitted with the analytic result \eqref{defectFinite properlength}.

\begin{figure}[t]
  \includegraphics[scale=0.45]{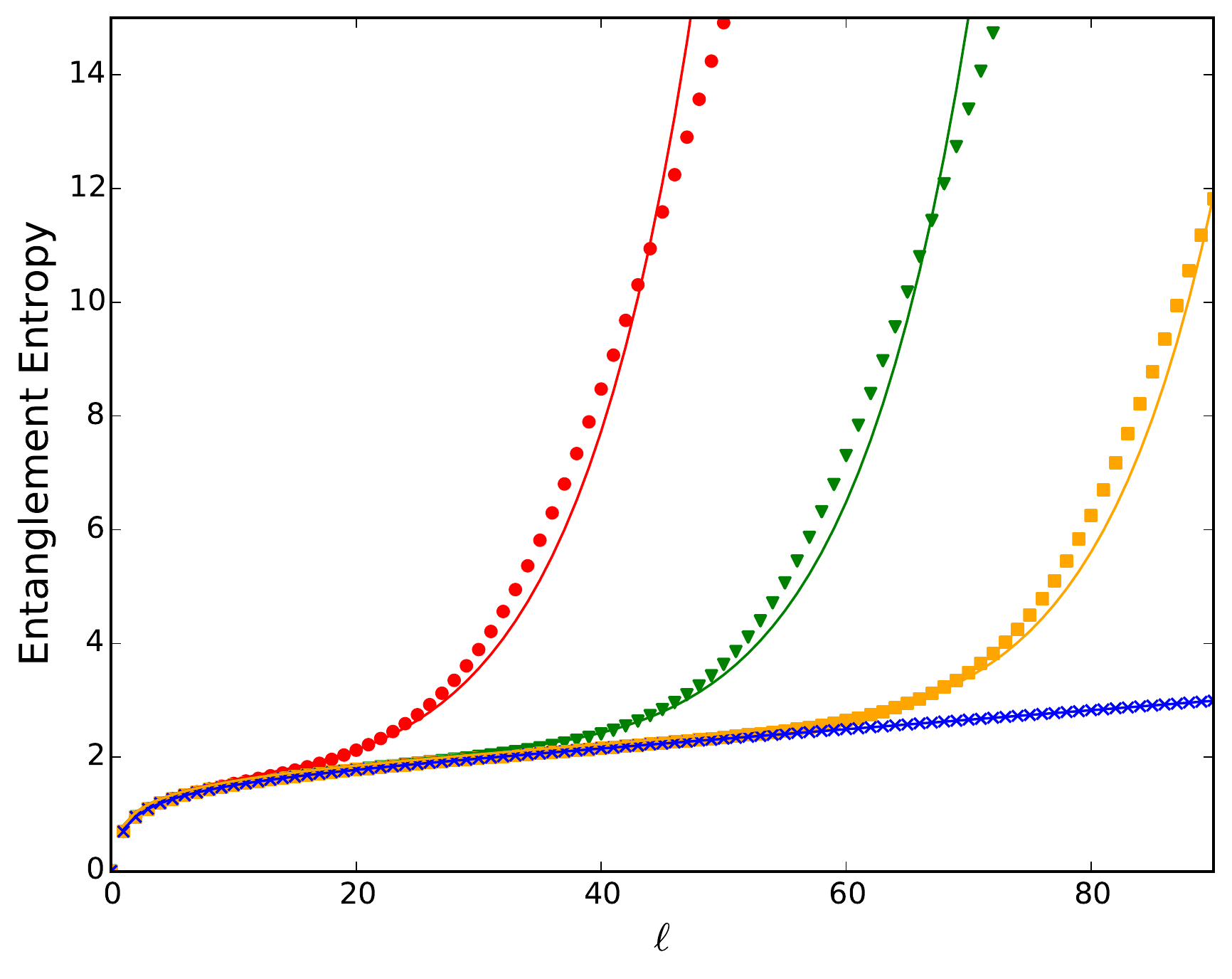}
  \caption{
    The numerically computed
    ``half-chain'' entanglement entropy
    % at $T=0$
    % for $h=0.5, 0.1, 0.05, 0.01$ from the top. 
    at finite temperature at $h=0.1$ 
    for $\beta=100, 1000, 10000, \infty$ from the top.
    The numerical data is fitted to Eq.\ \eqref{generalRainbowFiniteT},
    where we treat the non-universal constant part
    (independent of $\ell$ and $\beta$)
    as a fitting parameter.
  }
  \label{fig:halfChainPlot}
\end{figure}

%
%
%\begin{figure}[t]
%  \includegraphics[scale=0.4]{ent_rainbow_halfchain_vs_h.png}
%  \includegraphics[scale=0.4]{ent_rainbow_halfchain_finiteT.png}
%  \caption{
%    The numerically computed
%    ``half-chain'' entanglement entropy
%    (a) at $T=0$
%    for $h=0.5, 0.1, 0.05, 0.01$ from the top,
%    and (b) at finite temperature at $h=0.1$
%    for $\beta=100, 1000, 10000, \infty$ from the top.
%    The numerical data are fitted to  
%    $
%    S_A (\ell_\gamma;b, \eta_1)
%    = \frac{c}{3}
%    \log \left[ \frac{\beta}{\pi h \eta_1 e^{h\ell_{\gamma} /2}} \sinh \left( \frac{\pi\eta_1(e^{h\ell_{\gamma}} - 1)}{\beta} \right) \right] + b,
%    $
%where $\eta_1$ and $b$ are fitting parameters found to be $b=-0.51$ and
%$\eta_1=39.18$ for an interval anchored at an arbitrary point on one side of the
%defect.
%  }
%  \label{fig:halfChainPlot}
%\end{figure}

\subsection{Half-chain entanglement at zero temperature}

Let us now discuss the half-chain entanglement. 
For this purpose, it is convenient to use the global coordinate:  
\begin{align}
  ds^2_{AdS_3} =
  \frac{h^2 R^2}{\cos^2(h \Theta)}
  \left(d\Theta^2 + ds^2_{AdS_2} \right),
\end{align}
where $ds^2_{AdS_2}$ is the metric of ${\it AdS}_2$ which will be described below.
There are two asymptotic boundaries located at $\Theta=\pm \pi/2h$,
where two CFTs, one for each boundary, are defined.
%These CFTs are put on ${\it AdS}_2$ with the metric $ds^2_{{\it AdS}_2}$.
As for $ds^2_{{\it AdS}_2}$, it is also given in terms of the global coordinate
\begin{align}
  ds^2_{{\it AdS}_2}
  =
   \frac{1}{h^2\cos^2 \xi}
  (-d\tau^2 + d\xi^2), 
  \label{theta and xi}
\end{align}
% where $1/h$ is the radius of ${\it AdS}_2$,
$\tau\in (-\infty, +\infty)$ and $\xi\in (-\pi/2, \pi/2)$.
There are two asymptotic boundaries at $\xi=\pm \pi/2$.
The two CFTs are connected at the boundaries of ${\it AdS}_2$.
% It is also convenient to use the Poincar\'e metric.
%\begin{align}
%  ds^2_{AdS_2} =
%  h^2
%  \frac{-dt^2+d\eta^2}{\eta^2}.
%\end{align}
% In this metric, there is a boundary at $\eta=0$.

% 
% To discuss the entanglement entropy, 
% we now consider the RT minimal area surface 
% $\Gamma$ (which is a geodesic in this case). 

Combining the coordinate transformation (\ref{rainCoord}) with (\ref{sHolo}), we obtain the half-chain entanglement entropy. We can express this in terms of the proper length via $\ell= \frac{1}{h}\int_{\eta_1}^{\eta_2} \frac{d\eta}{\eta} \to \eta_2=\eta_1 \exp(h \ell)$:
\begin{align}
    S_A(\ell)= \frac{c}{3} \log \left[ \frac{2}{ \epsilon  h } \sinh \left( \frac{h\ell}{2} \right) \right].
 \label{zeroTempHalfChain}
\end{align}
 The most striking feature about this zero temperature result is that it demonstrates a volume law growth for larger values of $\ell$, with $h$ acting as an effective temperature. This is consistent with results from previous work on the rainbow chain,\cite{2015JSMTE..06..002R,2016arXiv161108559R} where it was found that the rainbow phase strongly resembles a thermofield double state. 
%The result is plotted in Fig.~\ref{fig:rainbowGeoLens}.
%
%the length of the entangling surface between $\pm \xi_0$ is therefore
%\begin{align}
%    {\it Len} ( \Gamma ( \ell_\gamma ) ) = R \log \left[ \frac{2}{\epsilon h} \sinh \left( \frac{h \ell_\gamma}{2} \right) \right].
%    \label{zeroTempHalfChain}
%\end{align}
%The result is plotted in Fig.~\ref{fig:rainbowGeoLens}.
%The resulting entanglement entropy is
%\begin{align}
%    S_A(\ell_\gamma)= \frac{c}{3} \log \left[ \frac{2}{ \epsilon  h } \sinh \left( \frac{h\ell_{\gamma}}{2} \right) \right].
%\end{align}

%\begin{figure}[t]
%%	\includegraphics[scale=0.25]{LenGamma_2.pdf}
%%	\includegraphics[scale=0.25]{Lengamma_1.pdf}
%	\includegraphics[scale=0.6]{gamma_vs_Gamma.pdf}
%  \caption{
%   % (a) The geodesic length (\ref{geoHalfChain}) plotted against $\xi_0$,
%   % (b) ~\ref{rainbowBdyLen}, the length of the corresponding boundary
%   % interval, (c) and
%    The geodesic length 
%    (\ref{geoHalfChain}) plotted as a function of
%    the length of the corresponding boundary interval (\ref{rainbowBdyLen}).}
%  \label{fig:rainbowGeoLens}
%\end{figure}

\subsection{Half-chain entanglement at finite temperature}

We can apply the same procedure of cutting off the geodesics in the BTZ spacetime at $\epsilon_z= \eta_{1,2} h \epsilon $. Here, we consider a geodesic anchored at arbitrary $\Theta= \pi/2h$ boundary points $\eta_1$ and $\eta_2$, where $\eta_2 > \eta_1$. Combining these cutoffs with our finite temperature master formula (\ref{finiteTempMaster}) yields
\begin{align}
    S_A(\eta_1,\eta_2; \beta)= \frac{c}{3} \log \left[ \frac{\beta}{\epsilon \pi h \sqrt{\eta_1 \eta_2}} \sinh \left( \frac{\pi (\eta_2 - \eta_1)}{\beta} \right) \right].
\end{align}
In terms of the proper length $\ell$, the half-chain entanglement is
\begin{align}
    &S_A(\ell; \beta, \eta_1) \nonumber \\
    &= \frac{c}{3} \log \left[ \frac{\beta}{\epsilon \pi h \eta_1 e^{h\ell /2}} \sinh \left( \frac{\pi \eta_1 (e^{h \ell} - 1)}{\beta} \right) \right].
    \label{generalRainbowFiniteT}
\end{align}
The low temperature $\beta \to  \infty$ limit
% of (\ref{generalRainbowFiniteT})
agrees with (\ref{zeroTempHalfChain}).

We have also computed numerical results for the half-chain entanglement in
Fig.~\ref{fig:halfChainPlot},
using the lattice model \eqref{numerH}
with the hopping amplitudes \eqref{envelope rainbow}.
There,
we choose two points $\eta_1$ and $\eta_2$, both located to the right of
the defect,
and set $\eta_1 = 12$ and increase $\ell$ by changing $\eta_2$. We use $\ell$ rather than the coordinates (\ref{properLenCoordTransform})
%, fitting the following function to the results:
%\begin{align}
%    S_A & (\ell_\gamma;b, \eta_1) \nonumber \\
%    &= \frac{c}{3}\left( \log \left[ \frac{\beta}{\pi h \eta_1 e^{h\ell_{\gamma} /2}} \sinh \left( \frac{\pi\eta_1(e^{h\ell_{\gamma}} - 1)}{\beta} \right) \right] + b \right),
%\end{align}
%where $\eta_1$ and $b$ are fitting parameters found to be $b=-1.53$ and
%$\eta_1=39.18$ for an interval anchored at an arbitrary point on one side of
%the defect.
We can see that the agreement between the numerical and holographic results is
excellent at low temperatures.
%Again, however, we begin to see a bit of deviation at higher temperatures for sufficiently long intervals, where our numerical results demonstrate volume law growth, and our holographic results do not.

%\begin{figure}[t]
%  \includegraphics[scale=0.3]{ee_half_T=0.pdf}
%  \includegraphics[scale=0.3]{ee_half_finiteT.pdf}
%  \caption{
%    (a) The numerically computed
%    ``half-chain'' entanglement entropy
%    at $T=0$
%    for $h=1, 0.5, 0.1$ from the top. 
%    (b) at finite temperature at $h=0.1$
%    for $\beta=10, 100, 1000, 10000$ from the top.
%  }
%\end{figure}

\section{Holographic dual of SSD}

\subsection{Zero temperature}

Now let us compute the holographic entanglement entropy of the SSD
at zero temperature using the metric \eqref{AdS SSD}.
Let us first consider an interval $[v_1, v_2]$ 
(where $0 \le v_1,v_2 \le 2 \pi$).
Cutting off at $u=\epsilon$ corresponds to a $v-$dependent $z$ cutoff of
$\epsilon_z= \frac{a\epsilon}{1-\cos(v)}$. 

Using (\ref{sHolo}) and the above cutoff, we find that the zero temperature entanglement entropy is
\begin{align}
  \label{SSD hol 2}
  S_A(v_1,v_2)
  &=\frac{c}{3} \log \left[
    \frac{2}{\epsilon}
    \sin \left(\frac{1}{2}\left|v_2-v_1\right| \right) \right].
\end{align}

The result
\eqref{SSD hol 2}
is just the entanglement entropy for a CFT on a finite length space with periodic boundary conditions and is what we expect from the SSD model.\cite{2011PhRvB..83f0414H}

The above holographic results can be readily reproduced 
by a CFT calculation. 
The metric on our conformal boundary ($u=0$) in imaginary time, $\tau$, is
\begin{align}
  ds^2
  %&= dv^2 + \frac{1}{a^2}(1-\cos(v))^2 d\tau^2
  %      \nonumber \\
      &= dv^2 + \frac{4}{a^2} \sin^4 \left( \frac{v}{2} \right)d\tau^2.
\end{align}
%
%Interestingly, this metric gives way to a well known time-evolution operator, the SSD, or sine-squared deformed Hamiltonian:
%\begin{align}
%H_{SSD}= \int dv \sin^2 \left( \frac{v}{2} \right) T_{\tau \tau}(\tau=0)
%\end{align}
%Let us again use the methods from our calculation with the dipole transformed
%boundary theory in order to compute the entanglement entropy of a CFT on this
%background metric.
%First, 
We factor out the $ \frac{4}{a^2}\sin^4 \left( \frac{v}{2} \right)$ and define a new variable 
\begin{align}
dx= \frac{a dv}{2 \sin^2 \left( \frac{v}{2} \right)} \longrightarrow 
x= - a \cot \left( \frac{v}{2} \right)
\end{align}
to write our metric as 
\begin{align}
ds^2 &= e^{2\sigma}ds_0^2,
       \quad
e^\sigma = \frac{2}{a} \sin^2 \left( \frac{v}{2} \right) = \frac{2a}{a^2+x^2},
\end{align}
where
$ds^2_0= dx^2 + d\tau^2$.
Since $x\in (-\infty, \infty)$, we can use the twist 2-point function on the full complex plane to compute the entanglement entropy \eqref{twist operator in flat space}.
% \begin{align}
%   \mathrm{Tr}\,(\rho^n_A)\big |_{ds^2_0}
%  %   &= \langle \mathcal{T}^{(n)}(x_1,t=0) \mathcal{T}^{(n)}(x_2,t=0)
%  %   \rangle_{ds^2_0} \nonumber \\
%   &= c_n \Big ( \frac{x_1-x_2}{\epsilon} \Big )^{-2\Delta^{(n)}}
% \end{align}
Then,
\begin{align}
  &
  \left. \mathrm{Tr}\,(\rho^n_A) \right|_{ds^2}
%    \nonumber \\
%  &= c_n e^{-\Delta^{(n)} [\sigma (u_1) + \sigma (u_2)]}
%    \left\langle
%    \sigma_{n}(x_1(v_1),0) \sigma_{-n}(x_2(v_2),0)
%    \right\rangle_{ds^2_0}
    \nonumber \\
  &=c_n e^{-\Delta^{(n)} [\sigma (u_1) + \sigma (u_2)]}
    \left[
    \frac{a \cot\left(\frac{v_2}{2} \right)-a \cot\left(\frac{v_1}{2} \right)}{\epsilon }\right]^{-2\Delta^{(n)}}
    \nonumber \\
  &=c_n \left[
    2 \sin\left( \frac{1}{2} \left( v_2-v_1 \right) \right )
    \right]^{-2\Delta^{(n)}}.
\end{align}
Taking the log and taking $n\rightarrow 1$,
we reproduce the holographic result \eqref{SSD hol 2}.

\subsection{Finite temperature}

We can once again use the cutoff $\epsilon_z=\frac{a\epsilon}{1-\cos(v)}$.
This time we will plug it into (\ref{finiteTempMaster}) in order to obtain finite temperature results for the SSD.
For the interval $[v_1,v_2]$, we obtain the following entanglement law:
\begin{align}
  &
  S_A( v_1,v_2; \beta)
  = \frac{c}{3}\log \left[ \frac{2}{ a } \sin \left( \frac{v_1}{2} \right)  \sin \left( \frac{v_2}{2} \right) \right]
  \nonumber\\
  & \qquad
    + \frac{c}{3} \log \left[ \frac{\beta}{\pi \epsilon}\sinh \left( \frac{\pi a}{ \beta} \frac{\sin \left(\frac{1}{2}(v_2-v_1) \right)}{ \sin \left( \frac{v_1}{2} \right)  \sin \left( \frac{v_2}{2} \right) } \right) \right] .
    \label{SSD hol 3}
\end{align}
We can confirm that at zero temperature,
$S_A(v_1,v_2; \beta \to \infty)$ agrees with
the result from the previous section \eqref{SSD hol 2}. 
Note, unlike the zero temperature result \eqref{SSD hol 2},
\eqref{SSD hol 3} depends on $a$.
%\textcolor{red}{
%At finite temperatures, however, we see the emergence of the length scale, $a$. This length scale will also appear in the numerics, interestingly enough.}
%$\beta \to \infty$, we have
%\begin{align}
%S_A(v_1,v_2; \beta \to \infty) \to \frac{c}{3}\log \left( 2\left| \frac{ \sin(\frac{1}{2}(v_2-v_1))}{\epsilon}  \right| \right),
%\end{align}
%which agrees with our zero temperature results from the previous section. 
If we center the interval about $\pi$,
so that $v_1= \pi -v_0$ and $v_2= \pi + v_0$ (where $v_0 \in [0, \pi )$),
\begin{align}
  &S_A( \pi-v_0,\pi+v_0; \beta)
  \nonumber \\
&= \frac{c}{3} \log \left[ \frac{2 \beta}{\pi a \epsilon} \cos ^2 \left( \frac{v_0}{2} \right) 
\sinh \left( \frac{2 \pi a}{\beta} \tan \left( \frac{v_0}{2} \right) \right) \right].
\end{align}

For comparison with the SSD model put on a line of finite
length $L$, 
we rescale 
$v_i \to x_i = L v_i/2\pi$
and
recall that the parameter $a$ and $L$ are related by \eqref{L vs a},
$a = {L}/{2\pi}$.
The entanglement entropy is then given by
\begin{align}
  &S_A(x_1,x_2; \beta)
    = \frac{c}{3}\log \left[ \frac{4\pi}{L}
    \sin \left( \frac{x_1}{\pi} \right)
    \sin \left( \frac{x_2}{\pi} \right) \right]
  \nonumber\\
  & \qquad
    + \frac{c}{3} \log \left[
    \frac{\beta}{\pi \epsilon}
    \sinh \left(
    \frac{L}{2 \beta}
    \frac{\sin \left(\frac{1}{\pi}(x_2-x_1) \right)}
    { \sin \left( \frac{x_1}{\pi} \right)
    \sin \left( \frac{x_2}{\pi} \right) } \right) \right].
\end{align}
%\textcolor{red}{(Check $v\to x$ is correct.)}
Once again,
if we center the interval about $\pi$,
so that $v_1= \pi -v_0$ and $v_2= \pi + v_0$ (where $v_0 \in [0, \pi )$),
\begin{align}
  &S_A( \pi-v_0,\pi+v_0; \beta)
  \nonumber \\
&= \frac{c}{3} \log \left[ \frac{4 \beta}{ L \epsilon} \cos ^2 \left( \frac{v_0}{2} \right) 
\sinh \left( \frac{L}{\beta} \tan \left( \frac{v_0}{2} \right) \right) \right].
                 \label{SSD result}
\end{align}

\begin{figure}[t]
  \includegraphics[scale=0.45]{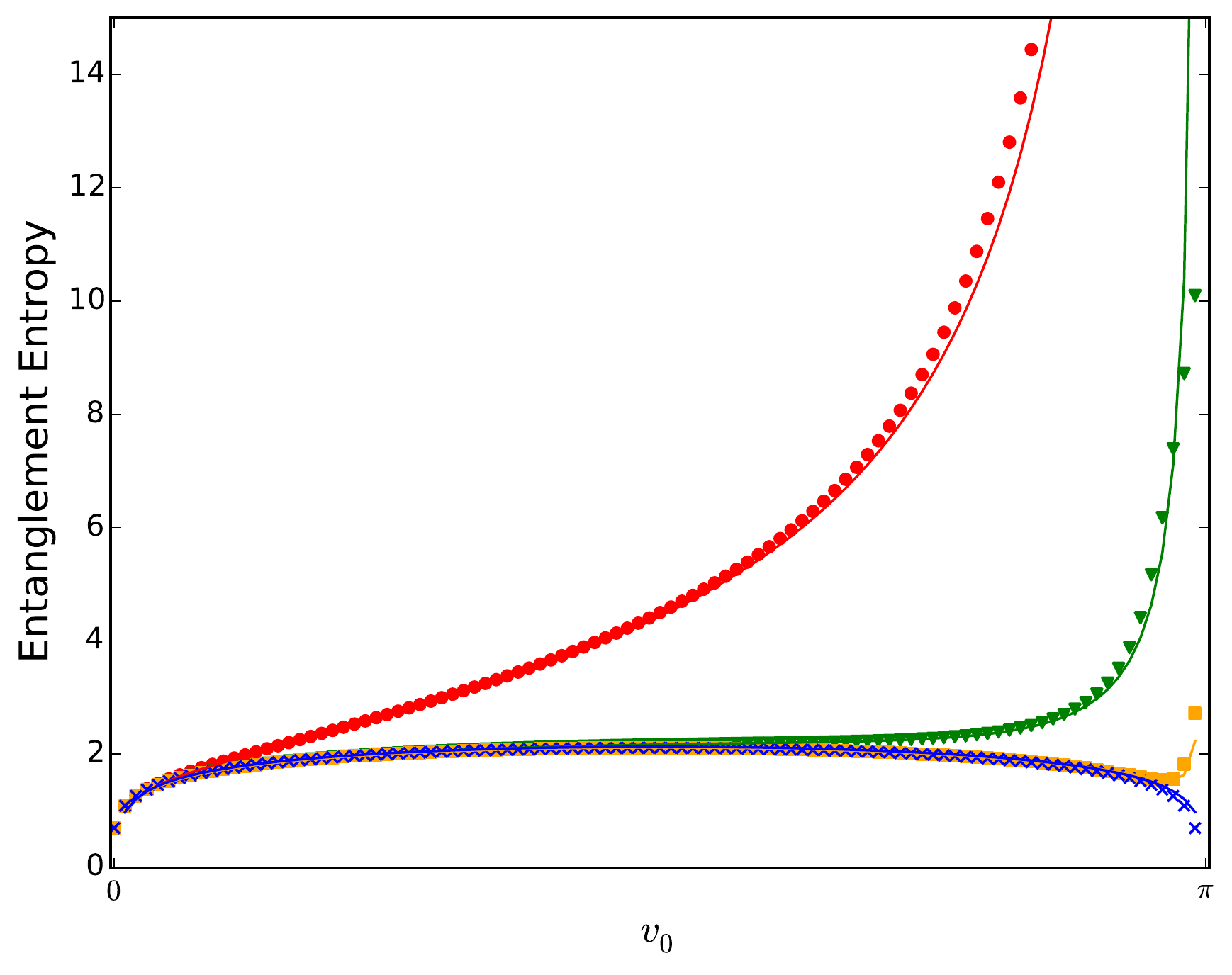}
  \caption{Finite temperature entanglement entropy
    in the SSD model,
    for an interval centered about $v=\pi$,
    and for $\beta= 10 ,100 ,1000, \infty$
    (from top to bottom).
    The dotted plots are numerical results from exact diagonalization and
    solid lines are the fitted analytic result \eqref{SSD result}
    where we treat the non-universal constant part
    (independent of $v_0$ and $\beta$) as a fitting parameter.
  }
  \label{fig:finiteSSD}
\end{figure}

In Fig.~\ref{fig:finiteSSD} we have plotted the entanglement entropy for
several values of $\beta$, computed numerically
from the Hamiltonian (\ref{numerH}) with
\begin{align}
  f_j= \frac{2}{a}
  \sin^2 \left(  \frac{j \pi}{N} \right)
  =
  \frac{4 \pi}{N-1}
  \sin^2 \left(  \frac{j \pi}{N} \right).
\end{align}
%at half filling.
Here, we put the over all factor
$2/a = 4\pi/(N-1)$
since the boundary metric
of our holographic setup is not
$ds^2_{{\it Mobius}}$ itself, but 
$ds^2_{{\it Mobius}}$ multiplied by the factor 
$(\cosh 2\gamma/a)^2 \to (2/a)^2$;
See Eq.\ \eqref{ds Mobius}.
%the nonuniversal length scale $a$ serves
% to stretch the radius of the $v$-coordinate circle or, equivalently,
The $N$-dependent multiplicative factor 
serves to compress the density of states of the SSD model.
This quantity only becomes relevant to the entanglement entropy at finite
temperature, as higher energy states become populated.

\subsection{Dipolar limit}

The dipolar limit of our holographic SSD model \eqref{AdS SSD dipolar}
can be studied analogously 
to the holographic duals of the rainbow and SSD models.

\paragraph{Zero temperature}

A geodesic in these coordinates
terminates at $u=0$, $v_0=-1/x_0$, where $x_0=x(z=0)$ is the anchor point of the
geodesic in the original Poincar\'e coordinates.
We introduce a UV cutoff $\epsilon$ in the radial, $u$-direction.
In terms of the boundary anchor point, the $z$ coordinate cutoff is 
\begin{align}
\epsilon_z=\frac{a\epsilon}{v_0^2+\epsilon^2} \approx \frac{a\epsilon}{v_0^2}.
\end{align}
Using (\ref{sHolo}) we have the following entanglement entropy:
\begin{align}
  S_A(v_0)=
  % \frac{l_{RT}}{4 G_N}=
  \frac{c}{3} \log\left ( \frac{2 v_0}{a\epsilon} \right ).
  \label{hol EE dipolar limit}
\end{align}
As we can see, the holographic entanglement entropy is unchanged by the dipolar
coordinate transformation in the bulk time slice.

The holographic result \eqref{hol EE dipolar limit}
can be readily reproduced from CFT calculations.  
From the previous subsection, we saw that the conformal boundary of the Poincar\'e metric in $u-v$ coordinates has the following metric:
\begin{align}
ds^2= -a^{-2}v^4dt^2 + dv^2.
\end{align}
Switching to imaginary time, $\tau$, and defining $x= {-1}/{v}$, or metric can be written as a Weyl-transformed flat metric in the $x-\tau$ coordinates,
$
ds^2= e^{2\sigma(x)}ds_0^2= e^{2\sigma(x)} d\bar{z}dz
$
where $z=x+i\tau$ and
\begin{align}
e^{2\sigma(x)}= a^{-2}v^4= \frac{1}{a^2x^4}.
\end{align}
%We want to calculate the 2-point function of twist operators located at two
%points $v_1$ and $v_2$ at a fixed time.
Using (\ref{cftCurved}) as before, we can compute the correlator on this particular curved background:
\begin{align}
  \mathrm{Tr}\, (\rho^n_A) |_{ds^2}
  &= c_n e^{-\Delta^{(n)} [\sigma (v_1) + \sigma (v_2)]}
    \left[ \frac{1}{a \epsilon v_1} - \frac{1}{a \epsilon v_2} \right]^{-2\Delta^{(n)}}
    \nonumber \\
%&= c_n(v_1v_2)^{-2\Delta^{(n)}}\left( \frac{1}{\epsilon v_1} - \frac{1}{\epsilon v_2} \right)^{-2\Delta^{(n)}} \nonumber \\
&=c_n \left[ \frac{a \epsilon}{v_2-v_1} \right]^{-2\Delta^{(n)}}.
\end{align}
Taking the log, multiplying by the appropriate factor,
and taking the replica limit $n\to 1$,
we recover \eqref{hol EE dipolar limit}.
%yields the $n^{th}$ Renyi entropy:
%\begin{align}
%S_A^{(n)}&= \frac{1}{1-n}\log \left( \mathrm{Tr}\, (\rho^n_A) |_{ds^2} \right)\nonumber \\
%%&= \frac{1}{1-n} \left[ \Delta^{(n)} \log \left( \frac{\epsilon}{v_2-v_1} \right) + \log(c_n) \right]\nonumber \\
%&=-\frac{c}{6} \left( \frac{n+1}{n} \right) \log \left( \frac{\epsilon}{v_2-v_1} \right) + \frac{\log(c_n)}{1-n}.
%\end{align}
%Taking $n\rightarrow 1$,
%we obtain the following Von Neumann entropy:
%\begin{align}
%S_A= \frac{c}{3}\log \left( \frac{v_2-v_1}{a\epsilon} \right),
%\end{align}
%up to an additive constant. Unsurprisingly, we get the same exact entanglement growth law as we had in the original field theory, as we saw in the holographic calculation.

\paragraph{Finite temperatures}

\begin{figure}[t]
\includegraphics[scale=0.45]{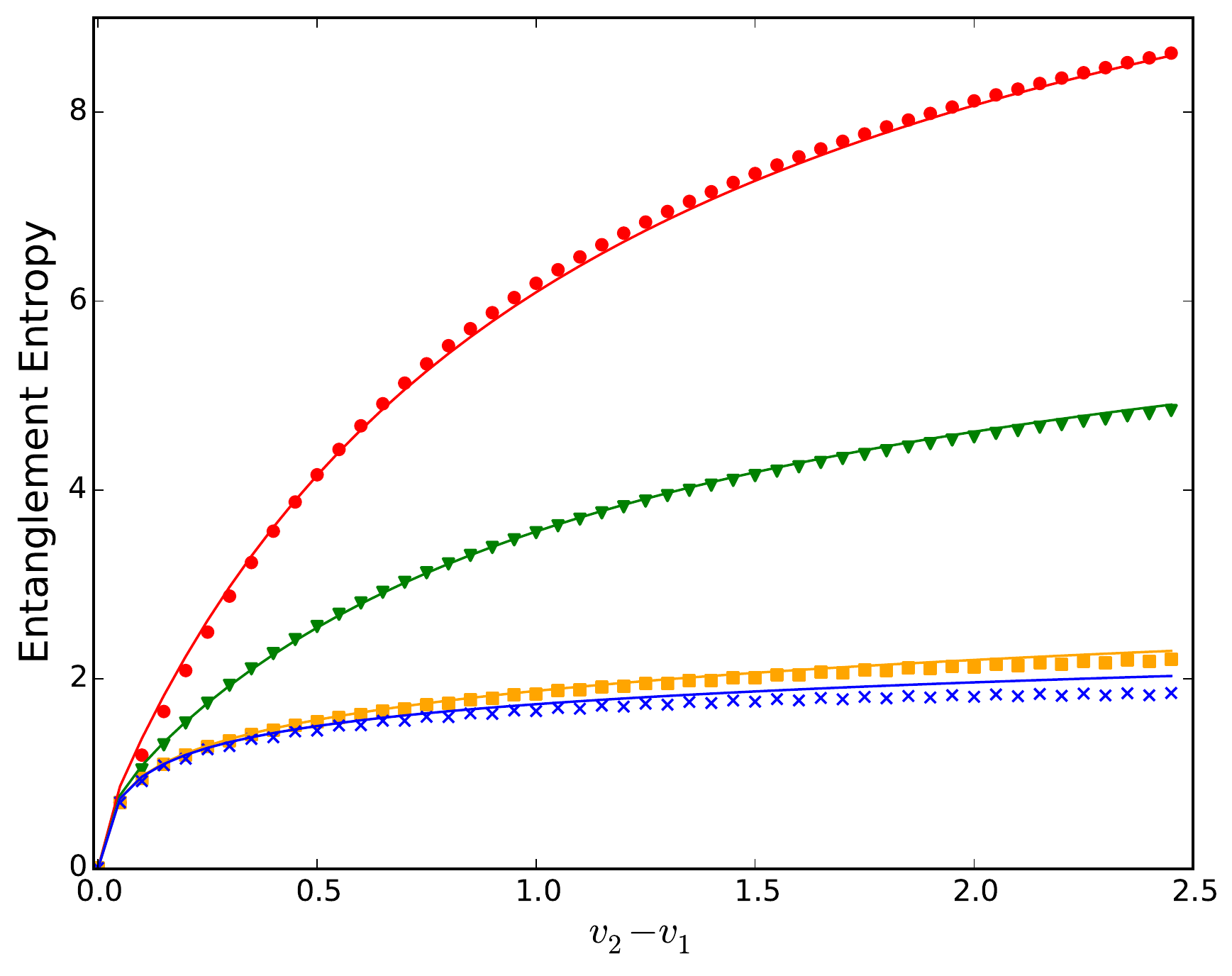}
\caption{Finite temperature entanglement entropy for the boundary theory of the
  dipole-foliated $AdS_3$ for an interval starting at $v=1.0$, for $\beta=
  1,2,10,\infty$ (from top to bottom).
  The plots are fit to \eqref{holoDipole}.
%  the following function:
%
%$  S_A  \left( v_1, v_2;  \beta,a, b \right)
%  = \frac{c}{3} \left( \log\left[\frac{\beta v_1 v_2}{2 \pi a}
%    \sinh \left( \frac{2\pi a}{\beta} \left( \frac{1}{v_2} - \frac{1}{v_1} \right) \right) \right] + b\right),
%$
where $v_1$ and $v_2$ are the left and right boundaries of the interval, respectively, $c$ is the central charge, and $b$ is a fitting parameter related to the lattice cutoff. This is just the small $v$ limit of the function in Fig. \ref{fig:finiteSSD}.
}
\label{fig:finiteDipole}
\end{figure}

As for the entanglement entropy at finite temperatures,
%As with the Rainbow chain, we can compute the entanglement entropy at a finite
%temperature.
plugging the $v_0$-dependent, constant $u$, cutoff, $\frac{\epsilon}{v_0^2}$ into (\ref{finiteTempMaster}), we have
\begin{align}
  S_A
  % = \frac{l_{RT}}{4 G_N}
  = \frac{c}{3}
  \log \left[ \frac{\beta v_0^2}{ \pi a \epsilon} \sinh\left(\frac{2 \pi a}{\beta v_0} \right) \right],
\end{align}
which at any temperature, $1/\beta$, asymptotes to the same logarithmic growth for large values of $v_0$. For generic endpoints, $v_1$ and $v_2$, we have
\begin{align}
\label{holoDipole}
  S_A
  %=  \frac{l_{RT}}{4 G_N}
  = \frac{c}{3} \log \left[
  \frac{\beta |v_1 v_2|}{2 \pi a \epsilon }
  \sinh\left(\frac{ 2 \pi a}{\beta }
  \left| \frac{v_2 - v_1}{v_1 v_2} \right| \right) \right].
\end{align}
In Fig.\ \ref{fig:finiteDipole},
we compare \eqref{holoDipole}
with numerical results obtained from
the Hamiltonian \eqref{numerH}
with $f_j = (10j/N-5)^2$.
%contains plots of the entanglement entropy
%calculated from the lattice Hamiltonian at half filling (dotted lines).
%The holographic formula \eqref{holoDipole} fits the numerical results remarkably well.
%It should be noted, however, that when $v_1$, the left boundary of our interval,
%becomes sufficiently small,
%the holographic results begin to disagree with the lattice calculations.
%This is unsurprising, as the BTZ space-time is not asymptotically $AdS$ if we
%take $u \to 0$ and $v \to 0$,
%and our interval will approach the black hole horizon.
%In the BTZ spacetime, we have $z<z_H$.
%In the dipole coordinates, $z = \frac{u}{u^2 + v^2}$, so that even at a fixed
%$u=\epsilon$, for sufficiently small $v$ our conformal boundary approaches the
%black hole horizon.
%Indeed, at $v= \sqrt{\epsilon/z_H}$, our conformal boundary actually touches the horizon. Thus, our dipole foliation breaks down at finite temperatures for boundary intervals near $v=0$.

\section{Conclusion}

In this paper, we combined
field theory, holographic, and numerical approaches, 
to investigate the entanglement entropy scaling
for (1+1)d CFTs put on various inhomogeneous backgrounds.  
At both zero and finite temperatures, we confirmed that 
all of these approaches deliver consistent results. 

While we focused on bipartite entanglement entropy,
the (holographic) approach laid out in this work 
can be used to compute other, related quantities, 
such as mutual information and entanglement negativity. 
Contrary to bipartite entanglement entropy in CFTs,
which is universal in the sense that it depends only on the central charge,
these quantities depend on details (precise operator content)
of CFTs, and can be used, e.g., to characterize 
quantum information scrambling. 
Negativity, in particular, could provide insight by drawing a clearer
distinction between genuine quantum correlation and classical, thermal entropy
in these curved systems; something that entanglement entropy misses.
We are currently investigating negativity in
(1+1)d CFTs on inhomogeneous spaces.

Other interesting extensions of this work could involve
applying our prescription to a wider varieties of inhomogeneous systems,
and to far-from-non-equilibrium systems,
such as systems undergoing quantum quenches~\cite{2018PhRvB..97r4309W}
and Floquet systems.~\cite{2018arXiv180500031W}

\acknowledgments
We thank Alexander Abanov, Jonah Kudler-Flam, Esperanza Lopez,
Gautam Satishchandran, Hassan Shapourian, Tsukasa Tada, and Xueda Wen for useful discussions. SR is supported by a Simons Investigator Grant from the Simons Foundation. This work was supported in part by the National Science Foundation grant DMR 1455296.

\appendix

\section{Application: Particles in a Potential Well}

%\textcolor{red}{(Update this paragraph.)}

\begin{figure}[t]
\includegraphics[scale=0.4]{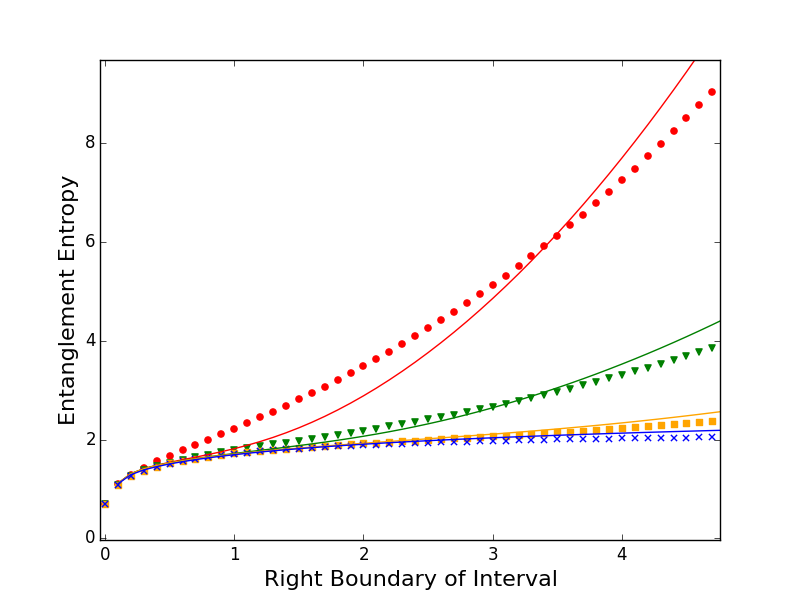}
\caption{Numerically computed entanglement entropy for free fermions in a Lorentzian potential well plotted at various temperatures, $\beta= 10$, $30$, $100$, $\infty$ plotted in dotted lines. (\ref{potentialEnt}) Plotted for the same temperatures in continuous lines.}
\label{fig:finiteLorentzian}
\end{figure}

The entanglement properties of nonrelativistic free fermions trapped in a potential well $V(x)$ can be described by a $c=1$ CFT on the following curved metric:~\cite{2016arXiv160604401D}
\begin{align}
    ds^2= dx^2 - v^2(x)dt^2
    \label{potentialMet}
\end{align}
where $v(x)= \langle \rho(x) \rangle= \sqrt{\frac{2}{m}( \mu - V(x))}$. Here, as in Refs. \onlinecite{2016arXiv160604401D} and \onlinecite{2017arXiv170102248B}, we have used the ``local density approximation". This is a semiclassical approximation in which the potential couples directly to the energy density, allowing for an easy application of our methods. Once we have this metric, we are implicitly using this approximation, and are no longer dealing directly with the nonrelativistic fermionic system. Henceforth, we use (\ref{potentialMet}) as our starting point for numerical and holographic calculations. Our holographic results are therefore not compared directly with the trapped fermion system, only with free fermions on the curved background given by (\ref{potentialMet}).

Starting with the Poincar\'e metric on $AdS_3$,
$ds^2_{{\it AdS}_3}= \frac{1}{z^2}\left[dz^2 +dy^2 -dt^2 \right]$,
we can induce the metric (\ref{potentialMet}) on the conformal boundary with the appropriate coordinate transformation
\begin{align}
  z= \frac{u}{v(x)},
  \quad y=  \int^x \frac{dx'}{v(x')}.
\end{align}
To leading order near the $u=0$ conformal boundary, our Poincar\'e  metric takes the form
\begin{align}
    ds^2 = \frac{1}{u^2} \left[ du^2 + dx^2 - v(x)^2 
    dt^2 \right] + \mathcal{O}(1)
\end{align}
Consider the example of a Lorentzian potential well, $V(x)= \frac{-1}{1+(x/a)^2}$, where $a$ controls the width of the well. We will set $\mu=0$. A lower chemical potential would result in an average particle density, $\langle \rho(x) \rangle$ that is zero outside of some finite region, requiring us to use BCFT methods.~\cite{2017arXiv170102248B} (Note, $\mu$ is {\it not} the chemical potential in our relativistic fermion system; it is better thought of as an input parameter for our metric). Our new coordinates are thus defined by (setting $m=2$ for convenience)
\begin{align}
  z &= u \sqrt{1+(x/a)^2},
  \nonumber \\ 
  y
    %&= \int^x dx' \sqrt{1+(x'/a)^2} \nonumber\\
    &= \frac{1}{2} \left(x\sqrt{1+(x/a)^2} + a \sinh^{-1} \left(x/a \right)  \right) + const.
\end{align}
This coordinate transformation implies a radial bulk cutoff of $\epsilon_z= \epsilon \sqrt{1+(x/a)^2}$. The finite temperature entanglement entropy for an interval between $-x_0$ and $x_0$ is thus
\begin{align}
  &S_A=  \frac{c}{3} \log \left[
    \frac{\beta}{\pi \epsilon \sqrt{1+(x_0/a)^2}} \right] \nonumber \\
    & + \frac{c}{3} \log \left[
      \sinh \left( \frac{\pi}{\beta} \left( x_0 \sqrt{1+(x_0/a)^2} + a \sinh^{-1} \left(x_0/a \right) \right) \right) \right].
    \label{potentialEnt}
\end{align}
The zero temperature limit of (\ref{potentialEnt}) is
\begin{align}
    S_A= \frac{c}{3} \log \left( \frac{1}{\epsilon} \left( x_0  + \frac{a \sinh^{-1} \left(x_0/a \right)}{\sqrt{1+(x_0/a)^2}} \right) \right)
\end{align}
In Fig.~\ref{fig:finiteLorentzian} we have plotted (\ref{potentialEnt}) for several temperatures alongside numerical results for the Lorentzian well. The agreement is qualitatively quite good. The noticeable discrepancies are likely due to the limitations of the bulk transformation we have used, and the resulting $\mathcal{O}(1)$ contributions to the metric that we are ignoring. It should be noted once again that we have not directly simulated nonrelativistic fermions in a potential well; we have assumed that the metric (\ref{potentialMet}) is a valid desription of the physics in this potential, and simulated free fermions living on this background geometry.

\bibliographystyle{unsrt}
\bibliography{reference.bib}

\end{document}